\documentclass[
 twocolumn,
 superscriptaddress,
 footinbib,
 amsmath,
 amssymb,
 aps,
]{revtex4-2}

\usepackage{graphicx} 
\usepackage{dcolumn} 
\usepackage{bm} 

\usepackage{hyperref} 

\usepackage{ascmac}

\usepackage[linesnumbered, ruled, vlined]{algorithm2e}

\usepackage{xcolor}

\DeclareMathOperator*{\argmax}{arg\,max}

\usepackage{slashed}

\usepackage{tikz}
\usetikzlibrary{quantikz}

\allowdisplaybreaks

\usepackage{MnSymbol}

\usepackage{soul}

\makeatletter
\newcommand{\vast}{\bBigg@{3.0}}
\newcommand{\Vast}{\bBigg@{4.0}}

\makeatother

\usepackage{accents}
\newlength{\dhatheight}

\usepackage[whole]{bxcjkjatype}

\usepackage{bbold}

\usepackage{amsthm}
\theoremstyle{plain}

\theoremstyle{definition}

\theoremstyle{remark}

\usepackage[compat=1.1.0]{tikz-feynhand}

\usepackage{simpler-wick}

\usepackage{xr}
\makeatletter
\@addtoreset{equation}{section}
\makeatother

\begin{document}

\preprint{APS/123-QED}

\title{Vicsek Model Meets DBSCAN: Cluster Phases in the Vicsek Model}

\author{Hideyuki Miyahara}
\email{miyahara@ist.hokudai.ac.jp, hmiyahara512@gmail.com}

\affiliation{
	Graduate School of Information Science and Technology,
	Hokkaido University, Sapporo, Hokkaido 060-0814, Japan
}

\author{Hyu Yoneki}

\affiliation{
	Graduate School of Information Science and Technology,
	Hokkaido University, Sapporo, Hokkaido 060-0814, Japan
}

\author{Tsuyoshi Mizohata}

\affiliation{
	Graduate School of Information Science and Technology,
	Hokkaido University, Sapporo, Hokkaido 060-0814, Japan
}

\author{Vwani Roychowdhury}


\affiliation{
	Department of Electrical and Computer Engineering, \\
	Henry Samueli School of Engineering and Applied Science, \\
	University of California, Los Angeles, California 90095
}

\date{\today}

\begin{abstract}
	The Vicsek model, originally proposed to elucidate the dynamics of bird flocking, undergoes a phase transition concerning the absolute value of the mean velocity.
	While clusters of agents are readily observed through numerical simulations of the Vicsek model, there is a lack of qualitative studies. We examine the clustering structure of the Vicsek model by employing DBSCAN, a widely used density-based clustering algorithm in the machine learning field.
	We find that as the radius specifying the interaction of the Vicsek model increases for a fixed noise magnitude, the model undergoes a crossover in the number of clusters from $\mathcal{O} (N)$ to $\mathcal{O} (1)$, where $N$ represents the number of agents.
	Additionally, we also propose a new order parameter to characterize flocking within clusters by utilizing the results of DBSCAN.
	Then, we identify at least four phases of the Vicsek model by combining the order parameter proposed by Vicsek \textit{et al.} with the number of clusters.
	To underscore the novelty of DBSCAN, we conduct mean shift, which is an alternative density-based clustering algorithm, demonstrating its lack of a linear relationship between the numbers of agents and clusters.
	This fact implies that we cannot define an order parameter in a naive manner using mean shift.
\end{abstract}


\maketitle


\section{Introduction}

The dynamics of self-propelling agents, often referred to as active matter, have garnered significant attention since the discovery of the phase transition in the Vicsek model, originally proposed as a model of bird flocking~\cite{Vicsek_001, Cavagna_001, Cavagna_002, Ramaswamy_001, Ginelli_002, Heffern_001, Solon_001}.
Active matter encompasses a wide array of systems, ranging from animal flocks~\cite{Parraish_001} and sheep herds~\cite{Ginelli_001} to human and non-human crowds~\cite{Murakami_001}, bacterial colonies~\cite{Bonner_001}, molecular motors~\cite{Harada_001}, tissue dynamics~\cite{Angelini_001}, crowd dynamics~\cite{Helbing_001, Angelini_001}, collective motion in rare events~\cite{Keta_001}, collective motion in disordered systems~\cite{Keta_002}, and nonreciprocal systems~\cite{Fruchart_001, Hanai_001}.
Numerical simulations have been pivotal in investigating these systems thus far.
Moreover, alongside numerical studies, nonequilibrium statistical-mechanical approaches grounded in the Boltzmann equation and hydrodynamic principles have furthered our understanding of the nature of active matter and the dynamics of self-propelling agents~\cite{Bertin_001, Bertin_002, Marchetti_001, Fodor_001, Toner_001}.
Additionally, a scaling relation within active matter has been discussed~\cite{Brambati_001}.
Inspired by the principles of topological condensed matter physics, the exploration of the robust structure of active matter, known as topological active matter systems, is also a thriving area of research~\cite{Shankar_001}.

In addition to the Ising-type phase transition of the Vicsek model found in Ref.~\cite{Vicsek_001}, the cluster structures of active matter have been the focus of intensive study.
These structures encompass various systems, including active Brownian particles~\cite{Fily_001, Peruani_002, Peruani_004, Peruani_005, Huepe_001}, active Brownian rods~\cite{Peruani_001, Yang_003, Abkenar_001, Shi_001, Peruani_003}, active colloids~\cite{Marchetti_002}, active fluids~\cite{Tjhung_001}, active networks~\cite{Kohler_001}, driven filaments~\cite{Schaller_001}, and bacterial colonies~\cite{Zhang_002}.
Motility-induced phase separation (MIPS), a concept akin to clustering phases, has garnered considerable interest~\cite{Cates_001}.
Moreover, the interplay between MIPS and frustration demonstrates a rich variety of phases in active matter~\cite{Adorjani_001}.

Another significant trend accelerating research in active matter is machine learning (ML)~\cite{Carleo_001, Das-Sarma_001}.
Recent studies on the application of ML to active matter are discussed in Refs.~\cite{Guo_001, Chen_001, de-Koning_001, Bhaskar_001}.
Particularly, clustering algorithms hold special significance since clustering structures appear ubiquitously in science and engineering, including active matter.
Reference~\cite{scikit_learn_clustering_001} summarizes several clustering algorithms, both well-known ones and those recently introduced, such as $k$-means, the EM algorithm with the Gaussian mixture model~\cite{Bishop_001, Murphy_001, Murphy_002}, OPTICS~\cite{Ankerst_001}, agglomerative clustering, DBSCAN (density-based spatial clustering of applications with noise), HDBSCAN (hierarchical density-based spatial clustering of applications with noise)~\cite{McInnes_001}, and mean shift~\cite{Ester_001}.
Depending on the choice of clustering algorithms, estimates of clusters present in a given dataset can drastically change and often fail to capture the appropriate structure.
Recently, cluster phases composed of bubbles have been reported~\cite{Tjhung_001, Shi_002, Nakano_001}.
As a related work, the network structure of the Vicsek model is investigated~\cite{Baglietto_001}.
Furthermore, research on applying ML to active matter is still relatively limited~\cite{Lopez_001, Ginot_001, Ferdinandy_001}.

In this paper, we investigate the clustering structure of the Vicsek model by applying DBSCAN, a clustering algorithm inspired by topological properties of data sets~\cite{Ester_001}.
First, we establish a mathematical connection between the Vicsek model and DBSCAN by reformulating the Vicsek model as an overdamped Langevin equation.
Subsequently, we demonstrate that the cost function of DBSCAN mathematically resembles the potential function of the Vicsek model.
Note that we do not insist the rigorous equivalence between them.
We then present our numerical findings indicating that the Vicsek model displays crossovers in the number of clusters, as discerned by DBSCAN.
Specifically, we observe a crossover from $\mathcal{O}(N)$ to $\mathcal{O}(1)$ in the relationship between the number of clusters and the number of agents, while maintaining a fixed density.
This transition occurs as the radius, which governs agent interactions in the Vicsek model, is increased while keeping the noise magnitude fixed.
Moreover, \textit{the result of DBSCAN enables us to define an intra-cluster order parameter, which characterizes a newly discovered phase, flocking within clusters}.
Combining the above quantities with the original order parameter proposed by Vicsek \textit{et al.}~\cite{Vicsek_001} together, we identify multiple phases in the Vicsek model.
\textit{While Ref.~\cite{Huber_001} discusses the coexistence of isotropic, nematic, and polar phases, our study reveals the coexistence of the cluster phase and the Ising-type order within the Vicsek model}.
Furthermore, to elucidate the novelty of DBSCAN, we conduct clustering using mean shift, which is also a widely used clustering algorithm, and show that it does not show a linear relationship between the numbers of agents and clusters.
Due to the lack of the linear relationship, an appropriate order parameter cannot be defined in a naive manner because the ratio of the number of clusters to that of agents always take a trivial value in the thermodynamic limit.

The rest of this paper is organized as follows: In Sec.~\ref{main_sec_Vicsek_model_001_001}, we introduce the Vicsek model~\cite{Vicsek_001}.
In Secs.~\ref{main_sec_DBSCAN_001_001} and \ref{main_sec_mean_shift_001_001}, we explain DBSCAN~\cite{Ester_001} and mean shift~\cite{Comaniciu_001}, respectively.
In Sec.~\ref{main_sec_numerical_simulations_001_001}, we present numerical simulations of the Vicsek model and the results of DBSCAN.
Finally, Sec.~\ref{main_sec_conclusions_001_001} summarizes our findings and concludes this paper.

\section{Vicsek model} \label{main_sec_Vicsek_model_001_001}

We first introduce the Vicsek model~\cite{Vicsek_001}, which was proposed as a possible explanation for bird flocking and has been successful in describing the phase transition of agents forming a flock.
Next, we derive its continuous-time variant, also known as the overdamped Langevin equation, and the potential energy of the Vicsek model.
This potential energy is minimized via the overdamped Langevin equation of the Vicsek model.
The potential energy of the Vicsek model plays an important role in discussing the similarity between the Vicsek model and DBSCAN.
In Appendix~\ref{main_sec_table_variables_001_001}, we summarize variables defined here in a table.

\subsection{Definition}

Let us consider a two-dimensional system of $n_\mathrm{ag}$ self-propelling agents.
We denote, by $\bm{x}_i (t)$ and $\bm{v}_i (t)$, the position and velocity of agent $i$ at time $t$, respectively.
The direction of $\bm{v}_i (t)$ is computed as
\begin{align}
	\bm{v}_i (t) & = v_\mathrm{abs}
	\begin{bmatrix}
		\cos \theta_i (t) \\
		\sin \theta_i (t)
	\end{bmatrix}, \label{main_eq_Vicsek_model_001_002}
\end{align}
and
\begin{align}
	\theta_i (t + \Delta t) & = \frac{1}{N_{i, r_\mathrm{V}} (t)} \sum_{\substack{j:            \\ \| \bm{x}_i (t) - \bm{x}_j (t) \|_\mathrm{F} \le r_\mathrm{V}}} \theta_j (t) + \Xi_i (t), \label{main_eq_Vicsek_model_001_003} \\
	\Xi_i (t)               & \sim [- \eta / 2, \eta / 2], \label{main_eq_Vicsek_model_001_004}
\end{align}
where $N_{i, r_\mathrm{V}} (t)$ is the number of elements $j$ that satisfy $\| \bm{x}_i (t) - \bm{x}_j (t) \|_\mathrm{F} \le r_\mathrm{V}$, while the magnitude of the velocity of each agent is constant: $\| \bm{v}_i (t) \|_\mathrm{F} = v_\mathrm{abs}$ where $\| \cdot \|_\mathrm{F}$ is the Frobenius norm.
Then the position of agent $i$ is updated by
\begin{align}
	\bm{x}_i (t + \Delta t) & = \bm{x}_i (t) + \bm{v}_i (t) \Delta t. \label{main_eq_Vicsek_model_001_001}
\end{align}
Eq.~\eqref{main_eq_Vicsek_model_001_004} means $\Xi_i (t)$ is randomly sampled from the uniform distribution whose upper and lower bounds are $\eta / 2$ and $- \eta / 2$, respectively.
Note that Eq.~\eqref{main_eq_Vicsek_model_001_003} is often called the metric Vicsek model since the summation is taken over all the agents that satisfy the metric condition.
As a variant of the Vicsek model, the topological Vicsek model, in which each agent interacts with its nearest neighbors independently from the actual distances, is of great interest~\cite{Ginelli_002}.

\subsection{Overdamped Langevin equation for the Vicsek model}

For small $\Delta t$, we can use the following approximation:
\begin{align}
	\theta_i (t + \Delta t) & \approx \theta_i (t) + \frac{\mathrm{d}}{\mathrm{d}t} \theta_i (t) \cdot \Delta t. \label{main_eq_first_order_approximation_theta_001_001}
\end{align}

From Eq.~\eqref{main_eq_Vicsek_model_001_003} and \eqref{main_eq_first_order_approximation_theta_001_001}, we have
\begin{align}
	\frac{\mathrm{d}}{\mathrm{d}t} \theta_i (t) & = \frac{1}{N_{i, r_\mathrm{V}} (t) \Delta t} \sum_{\substack{j: \\ \| \bm{x}_i (t) - \bm{x}_j (t) \|_\mathrm{F} \le r_\mathrm{V}}} \theta_j (t) - \frac{\theta_i (t)}{\Delta t} + \xi (t). \label{main_eq_continuous_Vicsek_model_001_001}
\end{align}
where $\xi (t) \coloneqq \frac{\mathrm{d}}{\mathrm{d}t} \Xi (t)$.

Assuming $N_{i, r_\mathrm{V}} (t) = N_{j, r_\mathrm{V}} (t)$ for $i$ and $j$ that belongs to the same cluster, we can rewrite Eq.~\eqref{main_eq_continuous_Vicsek_model_001_001} in the following form:
\begin{align}
	\frac{\mathrm{d}}{\mathrm{d}t} \theta_i (t) & = - \frac{\partial}{\partial \theta_i} V_{r_\mathrm{V}} (\{ \theta_i \}_{i=1}^{n_\mathrm{ag}}) \bigg|_{\{ \theta_i \}_{i=1}^{n_\mathrm{ag}} = \{ \theta_i (t) \}_{i=1}^{n_\mathrm{ag}}} + \xi (t), \label{main_eq_continuous_Vicsek_002_001}
\end{align}
where
\begin{align}
	V_{r_\mathrm{V}} (\{ \theta_i \}_{i=1}^{n_\mathrm{ag}}) & \coloneqq - \frac{1}{2 \Delta t} \sum_{i=1}^{n_\mathrm{ag}} \frac{1}{N_{i, r_\mathrm{V}}} \theta_i \Vast( \sum_{\substack{j: \\ \| \bm{x}_i - \bm{x}_j \|_\mathrm{F} \le r_\mathrm{V}}} \theta_j \Vast) \nonumber \\
	                                                        & \quad + \frac{1}{2 \Delta t} \theta_i^2                                                                                      \\
	                                                        & = - \frac{1}{2} \sum_{i=1}^{n_\mathrm{ag}} \sum_{\substack{j:                                                                \\ \| \bm{x}_i - \bm{x}_j \|_\mathrm{F} \le r_\mathrm{V}}} \frac{\theta_i \theta_j}{N_{i, r_\mathrm{V}}}  + \frac{1}{2 \Delta t} \theta_i^2. \label{main_eq_Vicsek_potential_001_001}
\end{align}
In the Vicsek model, Eq.~\eqref{main_eq_Vicsek_potential_001_001} is minimized via Eq.~\eqref{main_eq_continuous_Vicsek_002_001}.
Note that the condition of $N_{i, r_\mathrm{V}} = N_{j, r_\mathrm{V}}$ for $i$ and $j$ that belongs to the same cluster does not hold rigorously in the Vicsek model and it implies that the interaction of the Vicsek model is not reciprocal.

\subsection{Order parameter of the Vicsek model}

In Ref.~\cite{Vicsek_001}, the following function was investigated as the order parameter:
\begin{align}
	v_\mathrm{op} (t) & \coloneqq \frac{1}{v_\mathrm{abs} n_\mathrm{ag}} \bigg\| \sum_{i = 1}^{n_\mathrm{ag}} \bm{v}_i (t) \bigg\|_\mathrm{F}. \label{main_eq_order_parameter_Vicsek_001_001}
\end{align}
Vicsek \textit{et al.} reported that Eq.~\eqref{main_eq_order_parameter_Vicsek_001_001} describes the phase transition of the Vicsek model.

\section{DBSCAN} \label{main_sec_DBSCAN_001_001}

To investigate the clustering structure of the Vicsek model, we apply DBSCAN to the Vicsek model.
We first review DBSCAN~\cite{Ester_001, Schubert_001, Gao_001, Kang_001, Harris_001}.
Then, we derive the cost function of DBSCAN.
To the best of our knowledge, the cost function of DBSCAN has not been explicitly stated in the literature.
We show a correspondence between the cost function of DBSCAN and the potential function derived for the Vicsek model in the previous section.
This correspondence also explains why we have chosen DBSCAN.
Note that the discussion shown in this section is not the main claim of this paper; rather its aim is to support the numerical simulations shown later.
At the end of this section, we introduce new order parameters to define new phases of the Vicsek model.
In Appendix~\ref{main_sec_table_variables_001_001}, we summarize variables defined here in a table.

\subsection{Algorithm}

Clustering is an important task in many fields, and many different algorithms have been proposed~\cite{scikit_learn_clustering_001}.
Among these, DBSCAN~\footnote{DBSCAN is the abbreviation for Density-Based Spatial Clustering of Applications with Noise.} is widely used~\cite{Ester_001, Schubert_001, Gao_001, Kang_001, Harris_001}.
In this section, we describe the problem setting and the novelties of this approach.
Let $D_\mathrm{all} \coloneqq \{ \bm{x}_i \}_{i=1}^{n_\mathrm{ag}}$ be the dataset comprising $n_\mathrm{ag}$ data points.
The problem is to estimate $\{ l_i \}_{i=1}^{n_\mathrm{ag}}$, where $l_i \in \{ 0, 1, 2, \dots, n_\mathrm{cl} \}$ is the label of the $i$-th data point, with $l_i \ge 1$ meaning that the $i$-th data point belongs to the $l_i$-th cluster, and $l_i=0$ meaning that the $i$-th data point is an outlier and does not belong to any of the clusters.
There are two key points in this problem setting.
The first is to estimate the number of clusters, $n_\mathrm{cl}$, which is not a given number as in the $k$-means algorithm.
The second is that data points may be outliers, denoted by $l_i=0$.
This feature allows one to estimate a robust clustering structure.

Before getting into the details of DBSCAN, let us introduce the parameters of DBSCAN and explain the underlying concept behind DBSCAN.
The first parameter is the radius $r_\mathrm{D}$ that defines the neighbors of each data point, and the second parameter is the minimum number of neighbors $n_\mathrm{min}$ that each core point must have.
Otherwise, data points are recognized as outliers.
Additionally, we must specify a metric function for DBSCAN.
In this paper, we focus on the Euclidean distance.
The fundamental concept of DBSCAN is to classify data points into three categories: core points, border points, and outliers.
Core points have $n_\mathrm{min}$ or more neighbors, border points are reachable from core points but have fewer neighbors than $n_\mathrm{min}$, and outliers are not reachable from any data point that belongs to a cluster.
Moreover, core points and border points that belong to a cluster must form a cluster of which the number of agents is $n_\mathrm{min}$ or more.

We now describe the details of DBSCAN.
At the beginning of DBSCAN, we create the sets of visited data points $D_\mathrm{vst}$ and outliers $D_\mathrm{out}$ and initialize the two sets to the empty set: $D_\mathrm{vst} = \emptyset$ and $D_\mathrm{out} = \emptyset$.
We also set $n_\mathrm{cl} = 0$, where $n_\mathrm{cl}$ is the number of clusters.
Then, we randomly pick a data point from $D_\mathrm{all} \backslash D_\mathrm{vst}$ and run the following main loop.
In the main loop, we first add the selected data point to $D_\mathrm{vst}$ and compute the $r_\mathrm{D}$-neighbors of the selected data point.
We refer to the set of $r_\mathrm{D}$-neighbors of the selected data point as $D_\mathrm{nbh}$.
If the number of elements in $D_\mathrm{nbh}$ is smaller than $n_\mathrm{min}$, then the data points in $D_\mathrm{nbh}$ are added to $D_\mathrm{out}$.
Otherwise, we create a new cluster, increment $n_\mathrm{cl}$ by one, add the data points in $D_\mathrm{nbh}$ to the new cluster, and expand $D_\mathrm{nbh}$ by adding neighbors of $D_\mathrm{nbh} \backslash \bm{x}$ as much as possible.
We repeat the above main loop until $D_\mathrm{all} \backslash D_\mathrm{vst} = \emptyset$.
The role of the main loop is to find a core point and expand a cluster associated with the core point as much as possible.
Note that the first conditional branch of the main loop distinguishes core points from reachable points and outliers.
In Algo.~\ref{main_algo_DBSCAN_001_001}, the pseudocode of DBSCAN is shown~\cite{Schubert_001, Gao_001, Kang_001}.
Therein, two subroutines are used.
In Algos.~\ref{main_algo_DBSCAN_001_002} and \ref{main_algo_DBSCAN_001_003}, $\mathrm{region\_query} (\bm{x}, r_\mathrm{D})$ and $\mathrm{expand\_cluster} (\bm{x}, \{ C_k \}_{k=1}^{n_\mathrm{cl}}, D_\mathrm{vst}, r_\mathrm{D}, n_\mathrm{min})$ are shown, respectively.
Note that, by definition, $\bm{x} \in \mathrm{region\_query} (\bm{x}, r_\mathrm{D})$ is always satisfied.
\begin{algorithm}[t]
	\DontPrintSemicolon
	\caption{DBSCAN.}
	\label{main_algo_DBSCAN_001_001}
	\SetKwInOut{Input}{Input}
	\SetKwInOut{Output}{Output}
	\Input{Data set $D_\mathrm{all} \coloneqq \{ \bm{x}_i \}_{i=1}^{n_\mathrm{ag}}$, radius $r_\mathrm{D}$, minimum number of nodes in any cluster $n_\mathrm{min}$}
	\Output{Label set $\{ l_i \}_{i=1}^{n_\mathrm{ag}}$}
	set $D_\mathrm{vst} = \emptyset$, $D_\mathrm{out} = \emptyset$, and $n_\mathrm{cl} = 0$ \;
	set $l_i = 0$ for $i = 1, 2, \dots, n_\mathrm{ag}$ \;
	\For{$\bm{x} \in D_\mathrm{all} \backslash D_\mathrm{vst}$}{
		$D_\mathrm{vst} \leftarrow D_\mathrm{vst} \cup \{ \bm{x} \}$ \;
		$D_\mathrm{nbh} = \mathrm{region\_query} (\bm{x}, r_\mathrm{D})$ \;
		\uIf{$| D_\mathrm{nbh} | < n_\mathrm{min}$}
		{
			$D_\mathrm{out} \leftarrow D_\mathrm{out} \cup D_\mathrm{nbh}$ \;
		}
		\Else
		{
			$n_\mathrm{cl} \leftarrow n_\mathrm{cl} + 1$ \;
			create $C_{n_\mathrm{cl}}$ \;
			$C_{n_\mathrm{cl}}, D_\mathrm{vst} = \mathrm{expand\_cluster} (\bm{x}, \{ C_k \}_{k=1}^{n_\mathrm{cl}}, D_\mathrm{vst}, r_\mathrm{D}, n_\mathrm{min})$ \;
		}
	}
	\For{$i = 1$ \KwTo $n_\mathrm{ag}$}{
		\For{$k = 1$ \KwTo $n_\mathrm{cl}$}{
			\uIf{$\bm{x}_i \in C_k$}
			{
				$l_i = k$ \;
			}
		}
	}
	\KwRet $\{ l_i \}_{i=1}^{n_\mathrm{ag}}$ \;
\end{algorithm}
\begin{algorithm}[t]
	\DontPrintSemicolon
	\caption{$\mathrm{region\_query} (\bm{x}, r_\mathrm{D})$.}
	\label{main_algo_DBSCAN_001_002}
	\SetKwProg{Fn}{Function}{:}{end}
	\Fn{$\mathrm{region\_query} (\bm{x}, r_\mathrm{D})$}{
	\KwRet $\{ \bm{x}' \}_{\| \bm{x} - \bm{x}' \|_\mathrm{F} \le r_\mathrm{D}}$ \;
	}
\end{algorithm}
\begin{algorithm}[t]
	\DontPrintSemicolon
	\caption{$\mathrm{expand\_cluster} (\bm{x}, \{ C_k \}_{k=1}^{n_\mathrm{cl}}, D_\mathrm{vst}, r_\mathrm{D}, n_\mathrm{min})$.}
	\label{main_algo_DBSCAN_001_003}
	\SetKwProg{Fn}{Function}{:}{end}
	\Fn{$\mathrm{expand\_cluster} (\bm{x}, \{ C_k \}_{k=1}^{n_\mathrm{cl}}, D_\mathrm{vst}, r_\mathrm{D}, n_\mathrm{min})$}{
		$C_{n_\mathrm{cl}} \leftarrow C_{n_\mathrm{cl}} \cup \{ \bm{x} \}$ \;
		$D_\mathrm{nbh} = \mathrm{region\_query} (\bm{x}, r_\mathrm{D})$ \;
		\For{$\bm{x}' \in D_\mathrm{nbh}$}{
			\uIf{$\bm{x}' \not\in D_\mathrm{vst}$}
			{
				$D_\mathrm{vst} \leftarrow D_\mathrm{vst} \cup \{ \bm{x}' \}$ \;
				$D_\mathrm{nbh}' = \mathrm{region\_query} (\bm{x}', r_\mathrm{D})$ \;
				\uIf{$| D_\mathrm{nbh}' | \ge n_\mathrm{min}$}
				{
					$D_\mathrm{nbh} \leftarrow D_\mathrm{nbh} \cup D_\mathrm{nbh}'$ \;
				}
			}
			\uIf{$\bm{x}' \not \in \bigcup_{k = 1, 2, \dots, n_\mathrm{cl}} C_k$}
			{
				$C_{n_\mathrm{cl}} \leftarrow C_{n_\mathrm{cl}} \cup \{ \bm{x}' \}$ \;
			}
		}
		\KwRet $C_{n_\mathrm{cl}}, D_\mathrm{vst}$ \;
	}
\end{algorithm}

\begin{figure}[t]
	\centering
	\includegraphics[scale=0.70]{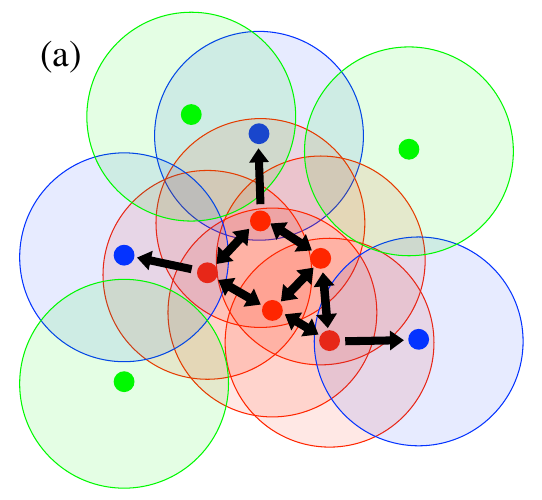}
	\includegraphics[scale=0.70]{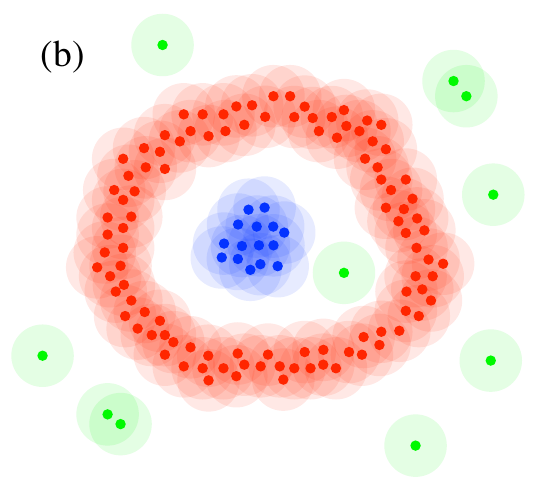}
	\caption{(Color online) Schematics of DBSCAN. (a) Red, blue, green points are core points, border points, and outliers, respectively. Red and blue points form one cluster. These definitions are consistent with setting $n_\mathrm{min}$ to any value in the set $ \{2, 3, 4\}$. (b) Red and blue data points form clusters and green data points are outliers. DBSCAN is applicable to linearly nonseparable datasets. These clustering results correspond to setting  $n_\mathrm{min}$ to any value in the set $\{3, 4, \cdots, 15\}$.}
	\label{main_shcematic_DBSCAN_001_001}
\end{figure}

In Fig.~\ref{main_shcematic_DBSCAN_001_001}, we show the schematic of DBSCAN.
In Fig.~\ref{main_shcematic_DBSCAN_001_001}(a), red, blue, and green points are core points, border points, and outliers, respectively.
We also set $n_\mathrm{min} \in (1, 4]$.
In Fig.~\ref{main_shcematic_DBSCAN_001_001}(b), red and blue points form clusters, and green points are outliers.
We also set $n_\mathrm{min}$ to any value from $ \{3, \ldots, 15\}$.
As shown in Fig.~\ref{main_shcematic_DBSCAN_001_001}(b), DBSCAN is applicable to linearly nonseparable datasets because it constructs a cluster by using a local structure of data points.

DBSCAN is usually applied to systems or datasets with the open boundary condition, but the Vicsek model has the periodic boundary condition.
With small modification, DBSCAN is applicable to a periodic boundary condition in Ref.~\cite{Turci_001}.

\subsection{Cost function of DBSCAN}

We first introduce the cost function of DBSCAN and discuss that DBSCAN yields solutions that are close to minima of this cost function, and that the DBSCAN algorithm implements an approximate version of a local update rule for minimizing the cost function.
We thus establish an optimization basis for DBSCAN.

Let $l_i \in \{0, 1, \ldots, N_\mathrm{cl} \}$ be the label assigned to the $i$-th data point $x_i$, where $N_\mathrm{cl}$ is an integer sufficiently larger than the actual number of clusters $n_\mathrm{cl}$ that DBSCAN would determine or present in the dataset.
Also, let $N_l$ be the number of data points that have been assigned the label $l$.
We then define a free energy type of the cost function:
\begin{align}
	F_\beta (\{ l_i \}_{i=1}^{n_\mathrm{ag}}) & \coloneqq U_{r_\mathrm{D}, n_\mathrm{min}} (\{ l_i \}_{i=1}^{n_\mathrm{ag}}) - \beta S (\{ l_i \}_{i=1}^{n_\mathrm{ag}}). \label{main_eq_cost_function_free_energy_DBSCAN_001_001}
\end{align}
The individual terms are then defined as:
\begin{align}
	U_{r_\mathrm{D}, n_\mathrm{min}} (\{ l_i \}_{i=1}^{n_\mathrm{ag}}) & \coloneqq - J \sum_{i=1}^{n_\mathrm{ag}} \delta_{l_i, \tilde{l}_i (r_\mathrm{D}, n_\mathrm{min})}, \label{main_eq_cost_function_energy_DBSCAN_001_001}                                                                 \\
	S (\{ l_i \}_{i=1}^{n_\mathrm{ag}})                                & \coloneqq - \bigg( \sum_{l=1}^{N_\mathrm{cl}} N_l \ln N_l \bigg) + \bigg( \sum_{l=1}^{N_\mathrm{cl}} N_l \bigg) \ln \bigg( \sum_{l=1}^{N_\mathrm{cl}} N_l \bigg), \label{main_eq_cost_function_entropy_DBSCAN_001_001}
\end{align}
where $\delta_{i, j}$ is the Kronecker delta function, which is unity for $i = j$ and zero otherwise,
\begin{align}
	 & \tilde{l}_i (r_\mathrm{D}, n_\mathrm{min}) \nonumber \\
	 & \quad \coloneqq
	\begin{dcases}
		\argmax_{l = 1, 2, \dots, n_\mathrm{cl}} N_{i, l} (r_\mathrm{D}, n_\mathrm{min}) & (\exists l, N_{i, l} (r_\mathrm{D}, n_\mathrm{min}) > 0), \label{main_eq_def_l_tilde_001_001} \\
		0                                                                                & (\text{otherwise}),
	\end{dcases}
\end{align}
and
\begin{align}
	N_{i, l} (r_\mathrm{D}, n_\mathrm{min}) & \coloneqq
	\begin{cases}
		\tilde{N}_{i, l} (r_\mathrm{D}) & (\tilde{N}_{i, l} (r_\mathrm{D}) \ge n_\mathrm{min}) \\
		0                               & (\text{otherwise}).
	\end{cases}, \label{main_eq_def_N_min_001_001}              \\
	\tilde{N}_{i, l} (r_\mathrm{D})         & \coloneqq \sum_{\substack{j = 1, 2, \dots, n_\mathrm{ag}: \\ \| \bm{x}_i - \bm{x}_j \|_\mathrm{F} \le r_\mathrm{D}}} \delta_{l, l_j}. \label{main_eq_def_N_tilde_001_001}
\end{align}
Furthermore, $\beta$ is a small real number $(0 < \beta \ll 1$), and $J > 0$ is a positive number.
Note that Eq.~\eqref{main_eq_cost_function_entropy_DBSCAN_001_001} is $n_\mathrm{ag}$ times the Shannon entropy with respect to the probability distribution of labels.

The choice of the cost function in Eq.~\eqref{main_eq_cost_function_free_energy_DBSCAN_001_001} and how it is related to the output of DBSCAN can be understood by explaining each term in it, and how a Monte Carlo optimization method can be applied to minimize the cost function.
Given any assignment of $\{ l_i \}_{i=1}^{n_\mathrm{ag}}$ to respective labels $\{ l \}_{l=1}^{N_\mathrm{cl}}$ a local update rule can be designed as follows.
First, for each data point $x_i$ and any given label $l$, we evaluate Eq.\eqref{main_eq_def_N_tilde_001_001} to get the number of data points within radius $r_\mathrm{D}$ from the $i$-th datapoint that also have the label $l$.
Then, Eq.~\eqref{main_eq_def_N_min_001_001} is used to discard the information of labels that do not satisfy the minimum number of datapoints in a cluster $n_\mathrm{min}$.
Then, we conduct the majority voting with respect to labels by Eq.~\eqref{main_eq_def_l_tilde_001_001}: that is, for each $x_i$ we find the label $\tilde{l}_i (r_\mathrm{D}, n_\mathrm{min})$ that has the maximum number of points assigned to it.
Finally, we update the label assignments by minimizing Eq.~\eqref{main_eq_cost_function_energy_DBSCAN_001_001} under the constraint of Eq.~\eqref{main_eq_cost_function_entropy_DBSCAN_001_001}.
A greedy strategy could be to assign the majority label as the new label of a datapoint.
However, without Eq.~\eqref{main_eq_cost_function_entropy_DBSCAN_001_001}, the number of clusters is more likely to be underestimated; thus Eq.~\eqref{main_eq_cost_function_entropy_DBSCAN_001_001} is necessary.
For instance, even when dealing with two well-separated clusters, combining them into a single cluster yields the same value for Eq.~\eqref{main_eq_cost_function_energy_DBSCAN_001_001}.

Returning to DBSCAN, the first and the dominant term of the cost function in Eq.~\eqref{main_eq_cost_function_energy_DBSCAN_001_001}, favors the same label of a node as obtained by a majority voting, as computed in Eq.~\eqref{main_eq_def_l_tilde_001_001}.
In this sense, the minimization of Eq.~\eqref{main_eq_cost_function_free_energy_DBSCAN_001_001} and DBSCAN have locally almost the same dynamics.
Moreover, since DBSCAN provides a method for starting a new cluster label when a new core node is encountered, it automatically incorporates the constraint in Eq.~\eqref{main_eq_cost_function_entropy_DBSCAN_001_001}.
Thus, any labeling assignment obtained by DBSCAN is already very close to a local minimum of the cost function.
That is, performing local updates for our cost function starting from a DBSCAN label assignment would keep it mostly unchanged.

\subsection{Why DBSCAN}

We have chosen DBSCAN to analyze the cluster phases of the Vicsek model although there have been many clustering algorithms proposed in the literature~\cite{scikit_learn_clustering_001}.
The simplest explanation is that its cost function, Eq.~\eqref{main_eq_cost_function_free_energy_DBSCAN_001_001}, shares similarities with the cost function of the Vicsek model, Eq.~\eqref{main_eq_Vicsek_potential_001_001}.
As mentioned in Ref.~\cite{Vicsek_001}, the Vicsek model becomes the random $XY$ model, in which agents interact with radius $r_\mathrm{V}$, in the limit $v_\mathrm{abs} \to 0$.
As shown in Eq.~\eqref{main_eq_cost_function_free_energy_DBSCAN_001_001}, datapoints within radius $r_\mathrm{V}$ also interact in DBSCAN.
Then, the cost function of DBSCAN can be seen as a discretized version of the cost function of the Vicsek model.
The continuous state variables $\{ \theta_i \}_{i=1}^{n_\mathrm{ag}}$ in the Vicsek model are replaced by the discrete label assignments $\{ l_i \}_{i=1}^{n_\mathrm{ag}}$ in DBSCAN.
Moreover in equation \ref{main_eq_Vicsek_potential_001_001}, instead of taking the average of $\{ \theta_i \}_{i=1}^{n_\mathrm{ag}}$ in a neighborhood $r_\mathrm{V}$, the majority of the labels of the datapoints in a neighborhood $r_\mathrm{V}$ is computed.
With these substitutions, the first term in Eq.\ref{main_eq_cost_function_energy_DBSCAN_001_001} is almost identical to \ref{main_eq_Vicsek_potential_001_001}.
Because the state variables are discrete in the DBSCAN formulation, the term Eq.~\eqref{main_eq_cost_function_entropy_DBSCAN_001_001} is necessary to prevent mode collapse.
However, their dynamical behaviors are different: the dynamics of DBSCAN is described by a greedy algorithm where only randomness is in the order in which datapoints are visited, while the dynamics of the Vicsek model is a stochastic process.
Additionally, in the case of the Vicsek model, each agent moves at a constant speed $v_\mathrm{abs}$; by taking the limit of $v_\mathrm{abs} \to 0$, their dynamical behaviors become more similar.

\subsection{Definition of new order parameters: cluster structure and intra-cluster order} \label{main_sec_new_order_parameter_001_001}

Introducing two parameters $\alpha$ and $\beta$, we consider the following relationship between $n_\mathrm{ag}$ (the total number of agents) and $n_\mathrm{cl}$ (the total number of clusters):
\begin{align}
	n_\mathrm{cl} & = \alpha + \beta n_\mathrm{ag}. \label{main_eq_relation_n_cl_n_ag_001_001}
\end{align}
Numerical simulations shown later indicates that $\alpha \in [0, 1]$ and $\beta \ge 0$, and we will see the following three regimes:
\begin{subequations}
	\begin{align}
		(\alpha = 0, \beta = 0) & : \text{no clusters},          \\
		(\alpha = 0, \beta > 0) & : \text{multiple clusters},    \\
		(\alpha = 1, \beta = 0) & : \text{single giant cluster}.
	\end{align}
\end{subequations}
More interestingly, we will observe the following ``crossovers":
\begin{subequations}
	\begin{align}
		(\alpha = 0, \beta = 0) & \Leftrightarrow (\alpha = 0, \beta > 0), \\
		(\alpha = 0, \beta > 0) & \Leftrightarrow (\alpha = 1, \beta = 0).
	\end{align}
\end{subequations}
We will also see, via numerical simulations, that the phase of $(\alpha = 1, \beta > 0)$ does not exist.

\begin{figure}[t]
	\centering
	\includegraphics[scale=0.70]{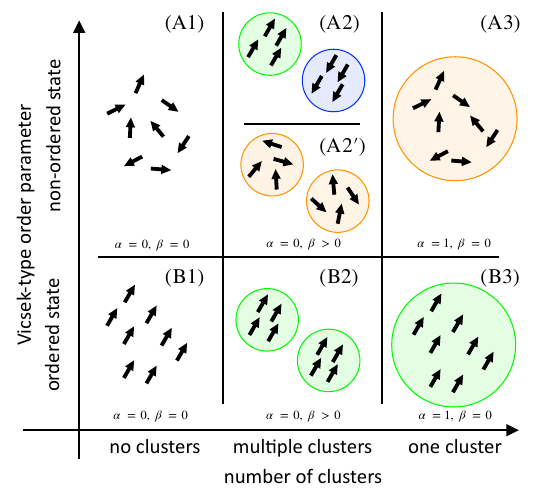}
	\caption{(Color online) Schematic of the phases of the Vicsek model: (A1) non-Vicsek-ordered state with no clusters, (A2, A2') non-Vicsek-ordered state with multiple clusters, (A3) non-Vicsek-ordered state with one cluster, (B1) Vicsek-ordered state with no clusters, (B2) Vicsek-ordered state with multiple clusters, and (B3) Vicsek-ordered state with one cluster. The difference between phase A2 and phase A2' is whether agents within clusters are Vicsek-ordered or not. \textit{Note that we call a state to be non-Vicsek-ordered} if the global average velocity is zero, even if each cluster may have ordered motion.}
	\label{main_schematic_DBSCAN_001_002}
\end{figure}

In Fig.~\ref{main_schematic_DBSCAN_001_002}, we present the schematic of the possible phase of the Vicsek model.
Therein we define new symbols: (A1) non-Vicsek-ordered state with no clusters, (A2, A2') non-Vicsek-ordered state with multiple clusters, (A3) non-Vicsek-ordered state with one cluster, (B1) Vicsek-ordered state with no clusters, (B2) Vicsek-ordered state with multiple clusters, and (B3) Vicsek-ordered state with one cluster~\footnote{We introduce the term Vicsek-ordered to emphasize the ordered state with respect to Eq.~\eqref{main_eq_order_parameter_Vicsek_001_001}.}.

To distinguish phases A2 and A2', we also define the following intra-cluster order parameter:
\begin{align}
	v_\mathrm{icop} (t) & \coloneqq \frac{1}{v_\mathrm{abs} n_\mathrm{ag}} \sum_{j=1}^{n_\mathrm{cl}} \bigg\| \sum_{i \in C_j} \bm{v}_i (t) \bigg\|_\mathrm{F}. \label{main_eq_intra-cluster_order_parameter_001_001}
\end{align}
The intra-cluster order parameter, Eq.~\eqref{main_eq_intra-cluster_order_parameter_001_001}, may allow us to distinguish phase A2 and phase A2'.

\section{Mean shift} \label{main_sec_mean_shift_001_001}

We here explain mean shift~\cite{Cheng_001, Comaniciu_001}.
In this paper, we mainly focus on DBSCAN as a clustering algorithm.
To elucidate the novelty of DBSCAN, we also conduct the numerical simulations of mean shift, which is also a density-based clustering algorithm.

In mean shift proposed by Cheng~\cite{Cheng_001}, we update $y_t$ by $y_{t+1} = m (y_t)$ with
\begin{align}
	m (y_t) & = \frac{\sum_{j=1}^N K (x_j; y_t, h) w (x_j) x_j}{\sum_{i=1}^N K (x_j; y_t, h) w (x_j)}. \label{main_eq_def_mean_shift_Cheng_001_001}
\end{align}
Here, $K (\cdot; \cdot, \cdot)$ is given by
\begin{align}
	K (x; \mu, h) & \coloneqq
	\begin{cases}
		1 & (\| x - \mu \| \le h), \\
		0 & (\text{otherwise}),
	\end{cases}
\end{align}
and $w (\cdot)$ is a nonnegative weight function.
The pseudocode of Cheng's mean shift is given in Algo.~\ref{main_algo_mean_shift_Cheng_001_001}.
\begin{algorithm}[t]
	\DontPrintSemicolon
	\caption{Cheng's mean shift.}
	\label{main_algo_mean_shift_Cheng_001_001}
	\SetKwInOut{Input}{Input}
	\SetKwInOut{Output}{Output}
	\Input{data set $\{ x_i \}_{i=1}^{n_\mathrm{ag}}$, input of the reference point $y$}
	\Output{output of the reference point $y_*$}
	set $t = 0$ and $y_0 \leftarrow y$ \;
	\While{until convergence}{
		compute $m (y_t)$ in Eq.~\eqref{main_eq_def_mean_shift_Cheng_001_001} \;
		$y_{t+1} \leftarrow m (y_t)$ \;
		$t \leftarrow t + 1$\;
	}
	set $y_* = y_T$ where $T$ is $t$ at convergence \;
	\KwRet $\{ y_* \}$ \;
\end{algorithm}
In Ref.~\cite{Cheng_001}, mode seeking is also discussed.

\section{Numerical simulations} \label{main_sec_numerical_simulations_001_001}

We first plot the snapshots of the Vicsek model with DBSCAN and mean shift to show the possibility of new phases in the Vicsek model.
Next, we draw the time-evolution of the number of clusters estimated by DBSCAN, the order parameter, and the intra-cluster order parameter.
Then, we show the phase diagram of the Vicsek model, in which we can see several different phases.
To confirm the well-definedness in the thermodynamic limit of the newly defined order parameters proposed in Sec.~\ref{main_sec_new_order_parameter_001_001}, we see the system-size dependence of the number of clusters.
Furthermore, we perform the regression to the results of DBSCAN and mean shift by assuming the following function:
\begin{align}
	f (x) & = a x^c + b, \label{main_eq_regression_function_001_001}
\end{align}
where $a, b$, and $c$ are the parameters to be estimated.
As motivated by Eq.~\eqref{main_eq_relation_n_cl_n_ag_001_001}, we also consider
\begin{align}
	\tilde{f} (x) & = \tilde{a} x + \tilde{b}, \label{main_eq_regression_function_002_001}
\end{align}
where $\tilde{a}$ and $\tilde{b}$ are the parameters to be estimated.
In addition, we also plot that of mean shift, and then see that it is not well-defined in the thermodynamic limit, that is, it is always zero in the thermodynamic limit.

\subsection{Snapshots}

\begin{figure}[t]
	\centering
	\includegraphics[scale=0.70]{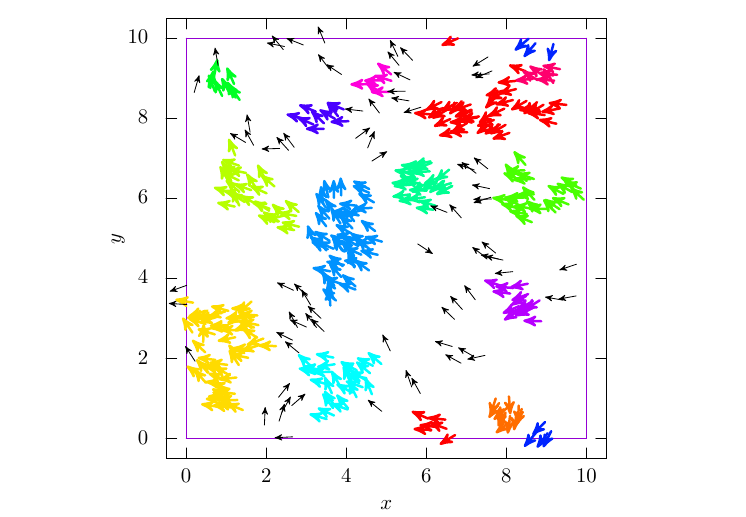}
	\includegraphics[scale=0.70]{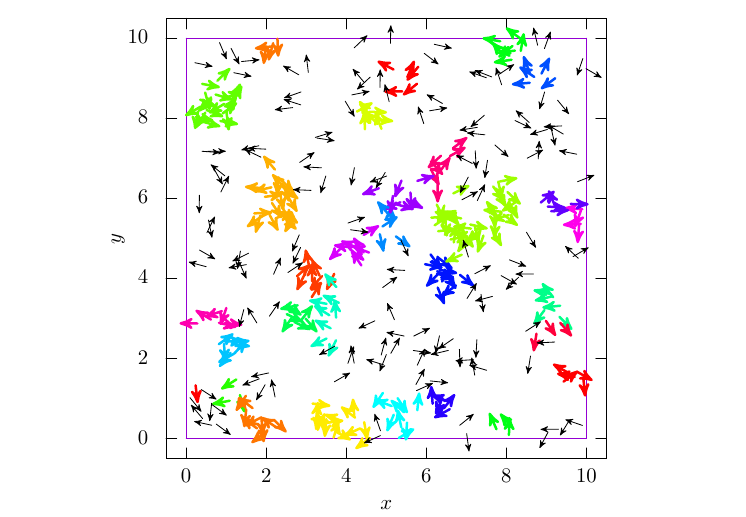}
	\caption{(Color online) Snapshots of the Vicsek model with DBSCAN for $r_\mathrm{V} = r_\mathrm{D} = 0.4$. We set (upper) $\eta = 1.0$ and (lower) $\eta = 4.0$. We also set $n_\mathrm{ag} = 400$, $L = 10.0$, $\Delta t = 1.0$, $v_\mathrm{abs} = 0.030$, $n_\mathrm{min} = 5$, and $t = 100$. Agents are colored based on their clusters and thin black arrows do not belong to any clusters. Here, $n_\mathrm{ag}$ is the total number of agents.}
	\label{main_fig_phase_diagram_snapshots_001_001}
\end{figure}

\begin{figure}[t]
	\centering
	\includegraphics[scale=0.70]{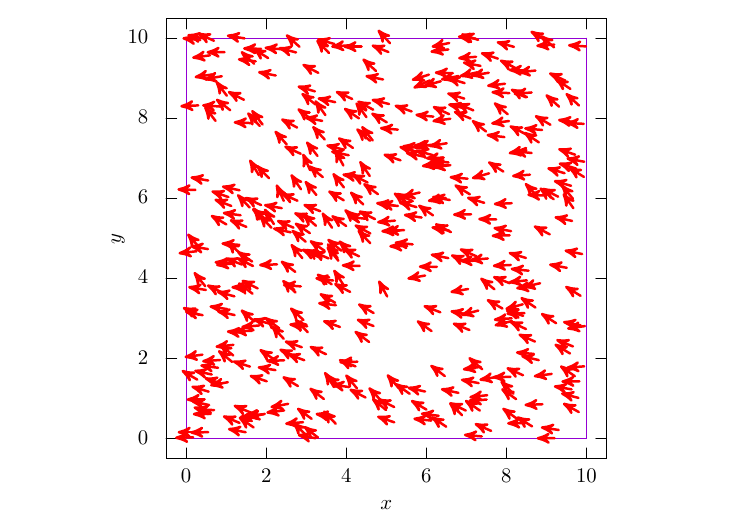}
	\includegraphics[scale=0.70]{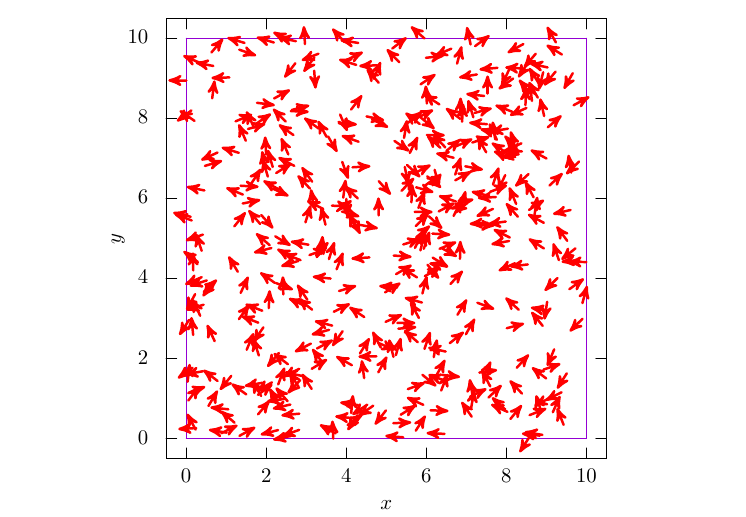}
	\caption{(Color online) Snapshots of the Vicsek model with DBSCAN for $r_\mathrm{V} = r_\mathrm{D} = 1.0$. We set (upper) $\eta = 1.0$ and (lower) $\eta = 4.0$. We also set $n_\mathrm{ag} = 400$, $L = 10.0$, $\Delta t = 1.0$, $v_\mathrm{abs} = 0.030$, $n_\mathrm{min} = 5$, and $t = 100$. Agents form a single cluster independently from the magnitude of $\eta$.}
	\label{main_fig_phase_diagram_snapshots_001_002}
\end{figure}

In Fig.~\ref{main_fig_phase_diagram_snapshots_001_001}, we display the snapshots of the Vicsek model with DBSCAN for $r_\mathrm{V} = r_\mathrm{D} = 0.4$.
In the figure, agents are colored based on clusters, and agents that do not belong to any cluster are depicted by thin black arrows.
When $\eta = 1.0$, there are many clusters, and agents are aligned in the same direction, whereas for $\eta = 4.0$, agents are not aligned in the same direction.
In Fig.~\ref{main_fig_phase_diagram_snapshots_001_002}, we present the snapshots of the Vicsek model with DBSCAN for $r_\mathrm{V} = r_\mathrm{D} = 1.0$.
In both cases, DBSCAN estimates a single large cluster because $r_\mathrm{D}$ is large.
For $\eta = 1.0$, the agents are aligned in the same direction, but for $\eta = 4.0$, the agents are directed almost randomly.

\begin{figure}[t]
	\centering
	\includegraphics[scale=0.70]{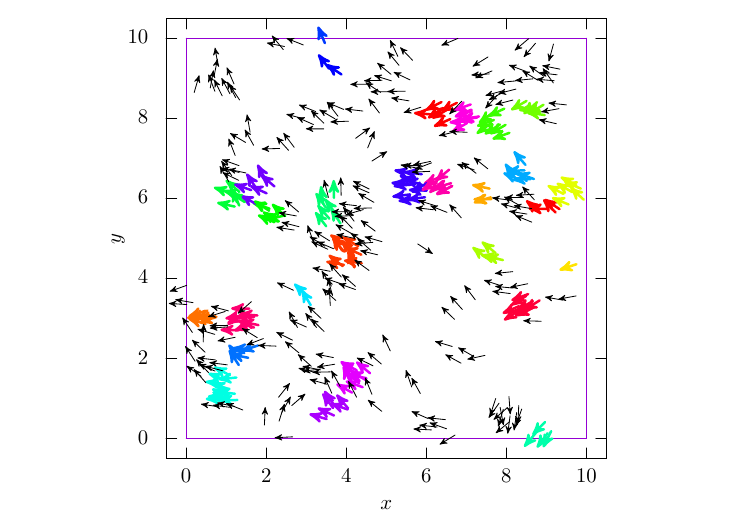}
	\includegraphics[scale=0.70]{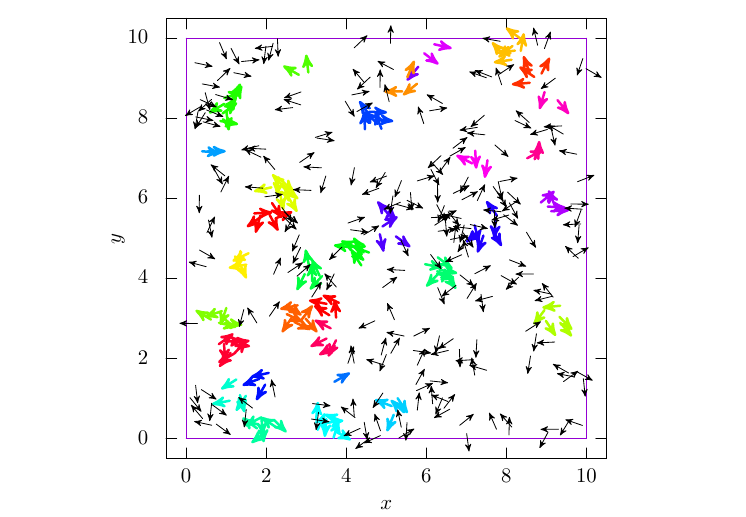}
	\caption{(Color online) Snapshots of the Vicsek model with mean shift for $r_\mathrm{V} = r_\mathrm{m} = 0.4$. We set (upper) $\eta = 1.0$ and (lower) $\eta = 4.0$. We also set $n_\mathrm{ag} = 400$, $L = 10.0$, $\Delta t = 1.0$, $v_\mathrm{abs} = 0.030$, $n_\mathrm{min} = 5$, and $t = 100$. The threshold is set to be $10 \%$. Agents are colored based on their clusters and thin black arrows belong to clusters in which the number of agents is less than $n_\mathrm{ag}$. Here, $n_\mathrm{ag}$ is the total number of agents.}
	\label{main_fig_phase_diagram_snapshots_002_001}
\end{figure}

\begin{figure}[t]
	\centering
	\includegraphics[scale=0.70]{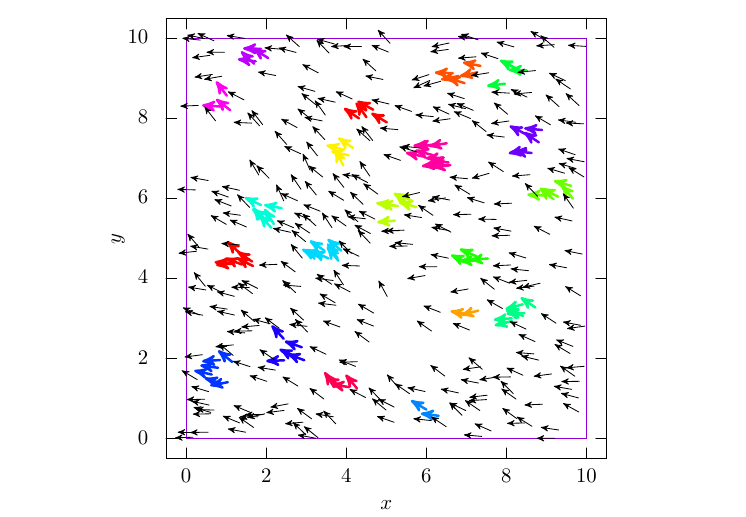}
	\includegraphics[scale=0.70]{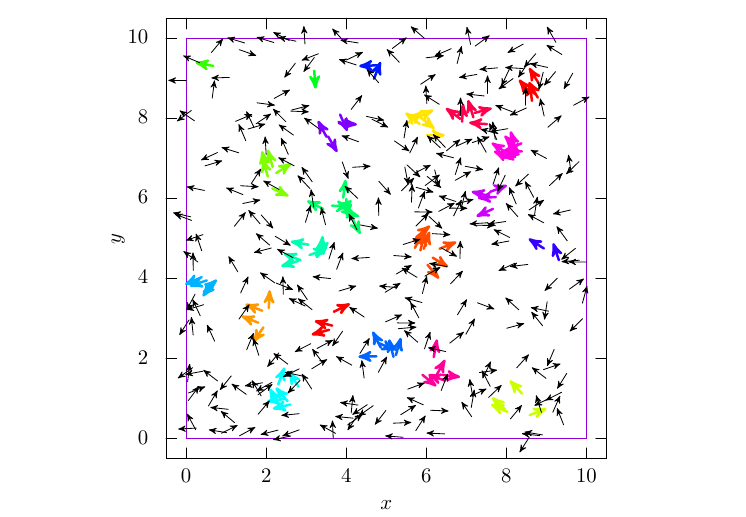}
	\caption{(Color online) Snapshots of the Vicsek model with mean shift for $r_\mathrm{V} = r_\mathrm{m} = 1.0$. We set (upper) $\eta = 1.0$ and (lower) $\eta = 4.0$. We also set $n_\mathrm{ag} = 400$, $L = 10.0$, $\Delta t = 1.0$, $v_\mathrm{abs} = 0.030$, $n_\mathrm{min} = 5$, and $t = 100$. The threshold is set to be $10 \%$. Agents are colored based on their clusters and thin black arrows belong to clusters in which the number of agents is less than $n_\mathrm{ag}$.}
	\label{main_fig_phase_diagram_snapshots_002_002}
\end{figure}

In Fig.~\ref{main_fig_phase_diagram_snapshots_002_001}, we display the snapshots of the Vicsek model with mean shift for $r_\mathrm{V} = r_\mathrm{m} = 0.4$.
The snapshot of the Vicsek model with mean shift for $r_\mathrm{V} = r_\mathrm{m} = 0.4$ and $\eta = 1.0$ shows that the size of the clusters is smaller compared to those obtained with DBSCAN.
In Fig.~\ref{main_fig_phase_diagram_snapshots_002_002}, we show the the snapshots of the Vicsek model with mean shift for $r_\mathrm{V} = r_\mathrm{m} = 1.0$.
The estimated number of clusters with mean shift is less dependent on $r_\mathrm{V}$.
We have ignored clusters in which the number of agents is less than $n_\mathrm{min}$ similarly to DBSCAN although mean shift does not originally have such a parameter.

In the remainder of this section, we quantify the differences among the aforementioned snapshots by computing macroscopic quantities.
As part of these macroscopic measures, we first compute the conventional order parameter proposed in Ref.~\cite{Vicsek_001}.
Additionally, we calculate the number of clusters and define a new order parameter within these clusters.

\subsection{Time-evolution of the estimated number of clusters, the order parameter, and the intra-cluster order parameter}

\begin{figure}[t]
	\centering
	\includegraphics[scale=0.70]{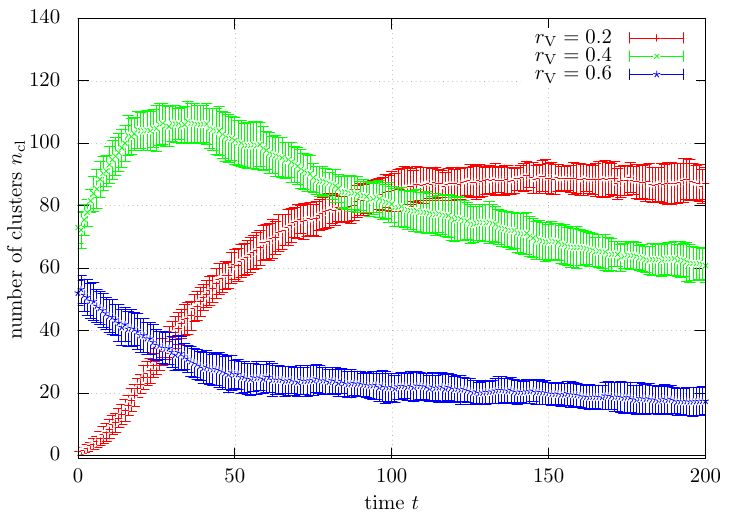}
	\includegraphics[scale=0.70]{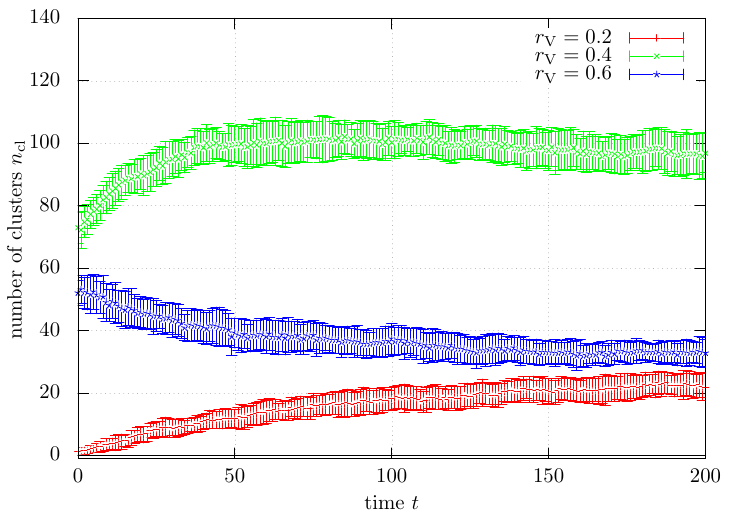}
	\caption{(Color online) Time-evolution of the estimated number of clusters with (top) $\eta = 2.0$ and (bottom) $\eta = 4.0$.}
	\label{main_fig_time_evolution_number_clusters_001_001}
\end{figure}

\begin{figure}[t]
	\centering
	\includegraphics[scale=0.70]{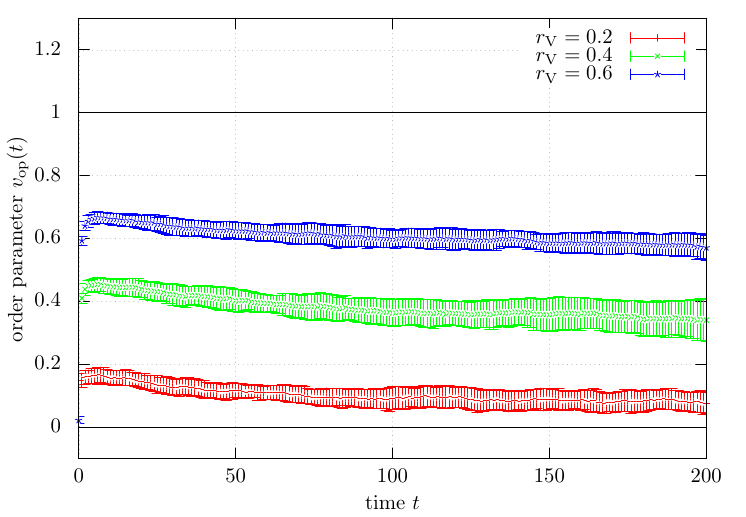}
	\includegraphics[scale=0.70]{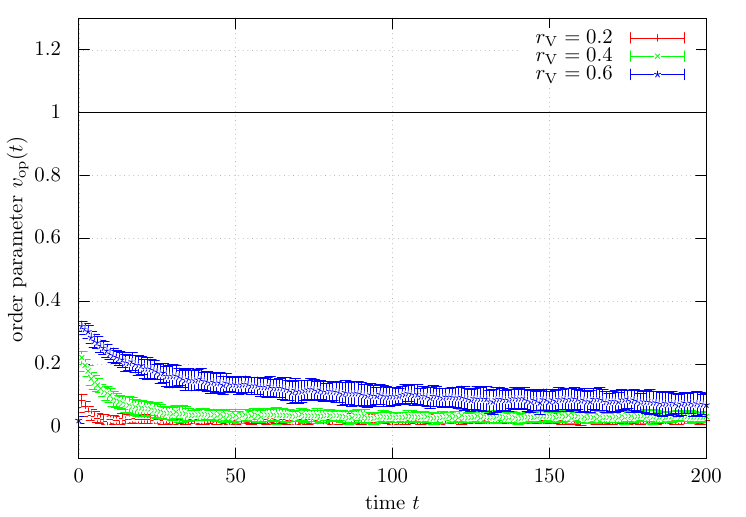}
	\caption{(Color online) Time-evolution of the order parameter with (top) $\eta = 2.0$ and (bottom) $\eta = 4.0$.}
	\label{main_fig_time_evolution_order_parameter_001_001}
\end{figure}

\begin{figure}[t]
	\centering
	\includegraphics[scale=0.70]{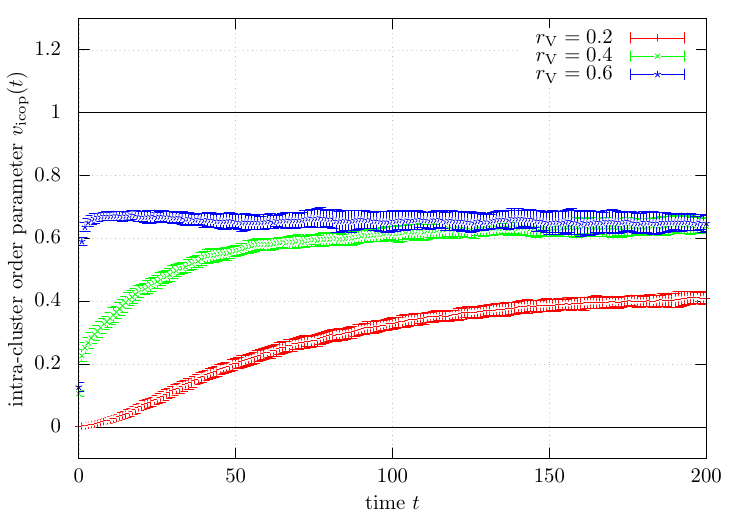}
	\includegraphics[scale=0.70]{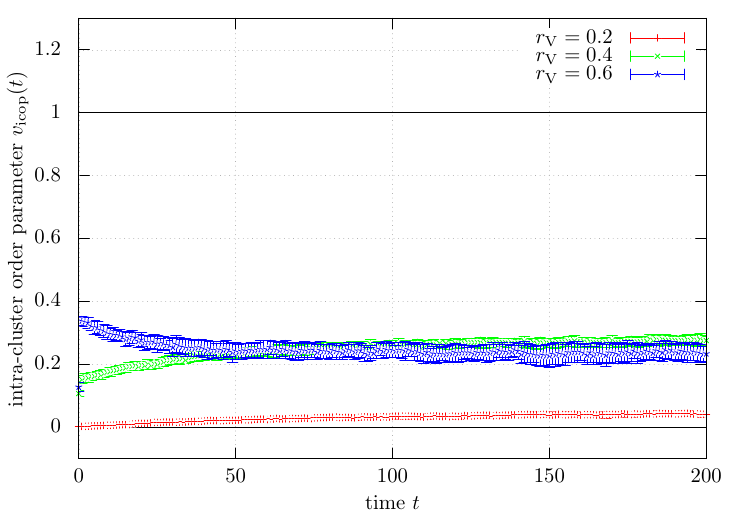}
	\caption{(Color online) Time-evolution of the intra-cluster order parameter with (top) $\eta = 2.0$ and (bottom) $\eta = 4.0$.}
	\label{main_fig_time_evolution_order_parameter_within_clusters_001_001}
\end{figure}

In Fig.~\ref{main_fig_time_evolution_number_clusters_001_001}, we show the time-evolution of the estimated number of clusters for several noise amplitudes.
For a wide range of the parameter regime, the number of clusters estimated by DBSCAN shows a fast relaxation.
The exception is the case of $r_\mathrm{V} = 0.4$ and $\eta = 2.0$, which is close to the critical point of flocking; so, its slow relaxation is reasonable.
In Fig.~\ref{main_fig_time_evolution_order_parameter_001_001}, we plot the time-evolution of the order parameter for several noise amplitudes.
The order parameter reaches its equilibrium at $t = 100$.
In Fig.~\ref{main_fig_time_evolution_order_parameter_within_clusters_001_001}, we show the time-evolution of the intra-cluster order parameter for several noise amplitudes.
The relaxation of the intra-cluster order parameter seems to be slower than that of the order parameter, but this result implies that $t = 100$ is sufficiently large.
Hereafter, we consider $t = 100$ for describing phase diagrams and $t = 100$ and $\eta = 4.0$ for plotting the system-size dependence of the estimated number of clusters, the order parameter, and the intra-cluster order parameter.

\subsection{Phase diagram and system-size dependence}

\begin{figure}[t]
	\centering
	\includegraphics[scale=0.70]{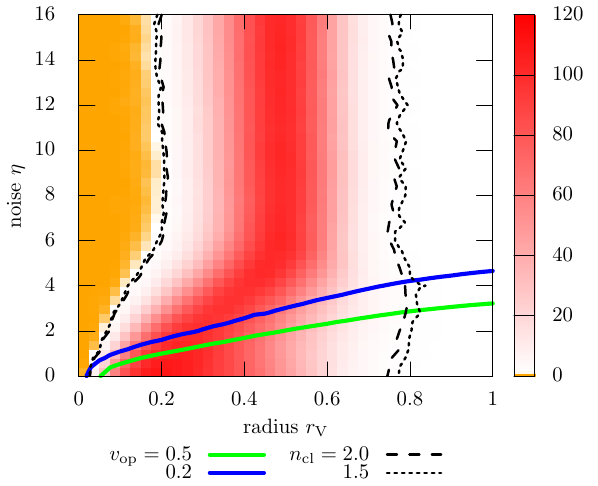}
	\caption{(Color online) Dependence of the number of clusters, $n_\mathrm{cl}$ (heat map), and $v_\mathrm{op} (t)$ (solid lines) on $r_\mathrm{V}$ (horizontal) and $\eta$ (vertical). We set $r_\mathrm{V} = r_\mathrm{D}$, $n_\mathrm{ag} = 1600$, $L = 20.0$, $\Delta t = 1.0$, $v_\mathrm{abs} = 0.030$, $n_\mathrm{min} = 5$, $t = 100$, and $n_\mathrm{rep} = 30$. The black dotted and dashed lines are contour lines for the number of estimated number of clusters. Here, $n_\mathrm{ag}$ is the total number of agents.}
	\label{main_fig_fidelity_001_003_001_001}
\end{figure}

In Fig.~\ref{main_fig_fidelity_001_003_001_001}, we plot the dependence of $n_\mathrm{cl}$ and $v_\mathrm{op} (t)$ on $r_\mathrm{V}$ and $\eta$.
We set $r_\mathrm{V} = r_\mathrm{D}$, $n_\mathrm{ag} = 1600$, $L = 20.0$, $\Delta t = 1.0$, $v_\mathrm{abs} = 0.030$, $n_\mathrm{min} = 5$, and $t = 100$.
We repeated the same calculations 30 times to compute the statistical means.
Hereafter we write the number of repetition $n_\mathrm{rep}$.
Figure~\ref{main_fig_fidelity_001_003_001_001} implies that the number of clusters $n_\mathrm{cl}$ exhibits crossovers twice upon increasing $r_\mathrm{V}$ with fixed $\eta$: from the orange regime to the red regime, and from the red regime to the white regime.
By combining the order parameter proposed in Ref.\cite{Vicsek_001}, Eq.~\eqref{main_eq_order_parameter_Vicsek_001_001}, with $n_\mathrm{cl}$, we identify at least five regimes, as the red and white regimes are each divided into two in addition to the orange regime.
More interestingly, in the regime of $\eta \ge 3.0$ and $r_\mathrm{V} \in [0.3, 0.5]$ of Fig.~\ref{main_fig_fidelity_001_003_001_001}, either phase A2 or phase A2' is expected to occur.
As we will see later, phase A2' does not exist in the Vicsek model.

\begin{figure}[t]
	\centering
	\includegraphics[scale=0.70]{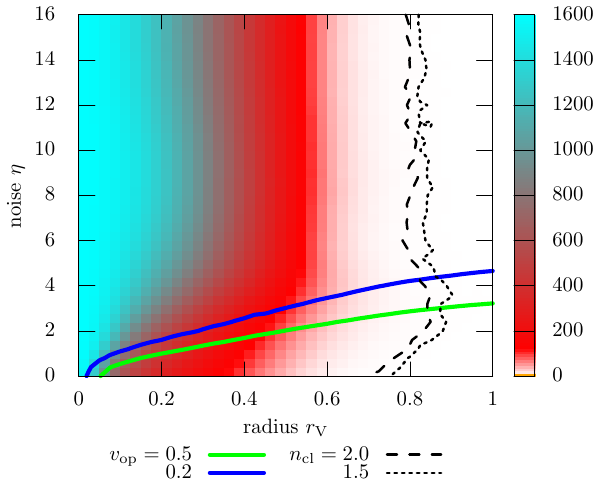}
	\caption{(Color online) Dependence of $n_\mathrm{cl}$ (heat map) and $v_\mathrm{op} (t)$ (lines) on $r_\mathrm{V}$ (horizontal) and $\eta$ (vertical). We set $r_\mathrm{V} = r_\mathrm{D}$, $n_\mathrm{ag} = 1600$, $L = 20.0$, $\Delta t = 1.0$, $v_\mathrm{abs} = 0.030$, $n_\mathrm{min} = 1$, $t = 100$, and $n_\mathrm{rep} = 30$. The black dotted and dashed lines are contour lines for the number of estimated number of clusters.}
	\label{main_fig_fidelity_001_003_001_002}
\end{figure}

To examine the robustness of the phases shown in Fig.~\ref{main_fig_fidelity_001_003_001_001} with respect to $n_\mathrm{min}$, we plot the dependence of $n_\mathrm{cl}$ and $v_\mathrm{op} (t)$ on $r_\mathrm{V}$ and $\eta$ for $n_\mathrm{min} = 1$ in Fig.~\ref{main_fig_fidelity_001_003_001_002}.
We use the same parameter values as in Fig.~\ref{main_fig_fidelity_001_003_001_001} for all other parameters.
The colorbars of Figs.~\ref{main_fig_fidelity_001_003_001_001} and \ref{main_fig_fidelity_001_003_001_002} are identical from 0 to 35.
The main difference between the two figures is the orange regime in Fig.~\ref{main_fig_fidelity_001_003_001_001} and the cyan regime in Fig.~\ref{main_fig_fidelity_001_003_001_002}.
This is because, in this regime, all the agents move almost randomly due to small $r_\mathrm{V}$ and large $\eta$.
In Fig.~\ref{main_fig_fidelity_001_003_001_001}, no clusters are found because of $n_\mathrm{min} = 5$, resulting in $n_\mathrm{cl} = 0$, whereas in Fig.~\ref{main_fig_fidelity_001_003_001_002}, all the agents are considered as clusters because of $n_\mathrm{min} = 1$, resulting in $n_\mathrm{cl} = 1600$.
Moreover, the red and white regimes in both figures are the same, regardless of the difference in $n_\mathrm{min}$.

\begin{figure}[t]
	\centering
	\includegraphics[scale=0.60]{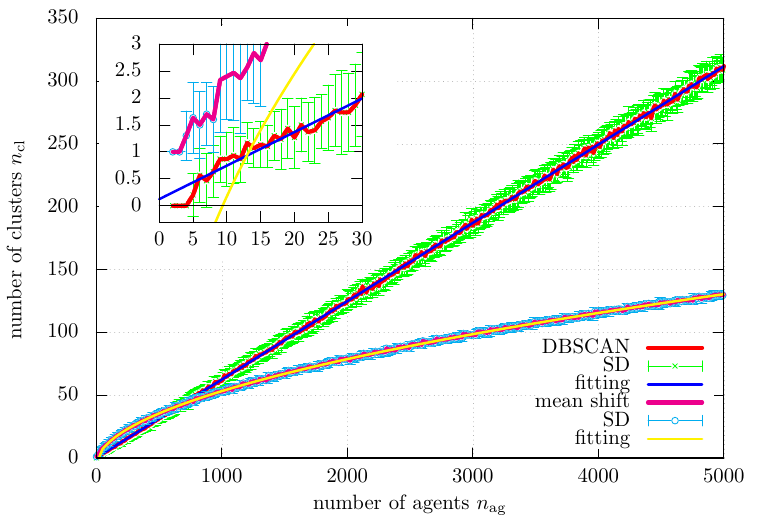}
	\caption{(Color online) Dependence of the expected value and standard deviation (SD) of $n_\mathrm{cl}$ on $n_\mathrm{ag}$. We set $r_\mathrm{V} = r_\mathrm{D} = 0.40$, $\eta = 4.0$, $\Delta t = 1.0$, $v_\mathrm{abs} = 0.030$, $n_\mathrm{min} = 5$, $t = 100$, and $n_\mathrm{rep} = 30$. We vary $L$ such that $n_\mathrm{ag} / L^2 = 4.0$. The regression functions, Eq.~\eqref{main_eq_regression_function_001_001}, for DBSCAN and mean shift are $f_\mathrm{D} (x) = (0.0623392 \pm 0.0004168) x^{(0.999958 \pm 0.0007858)} + (0.0898713 \pm 0.07635)$ and $f_\mathrm{m} (x) = (1.49398 \pm 0.01079) x^{(0.528996 \pm 0.0007938)} + (- 4.63302 \pm 0.09461)$, respectively. We used the data in the range of $n_\mathrm{ag} \in [10, 5000]$ to estimate the regression function.}
	\label{main_fig_fidelity_001_002_001_101}
\end{figure}

So far, we have discussed the phases for fixed $n_\mathrm{ag}$, but it is important to understand the behavior of the phases for large $n_\mathrm{ag}$, as we are interested in the thermodynamic limit.
In Fig.~\ref{main_fig_fidelity_001_002_001_101}, we plot the dependence of the expected value and standard deviation (SD) of $n_\mathrm{cl}$ on $n_\mathrm{ag}$~\footnote{We plot the standard deviation (SD) only when we plot the relationship between $n_\mathrm{cl}$ and $n_\mathrm{ag}$ since the standard error (SE) is too small in this scale. Note that SE is shown in error bars for other types of figures.}.
We set $r_\mathrm{V} = r_\mathrm{D} = 0.40$, $\eta = 4.0$, $\Delta t = 1.0$, $v_\mathrm{abs} = 0.030$, $n_\mathrm{min} = 5$, $t = 100$, and $n_\mathrm{rep} = 30$.
We vary $L$ such that $n_\mathrm{ag} / L^2 = 4.0$, i.e., we use parameters that belong to the red regime of Figs.~\ref{main_fig_fidelity_001_003_001_001} and \ref{main_fig_fidelity_001_003_001_002}.
Figure~\ref{main_fig_fidelity_001_002_001_101} clearly shows a linear relationship between $n_\mathrm{ag}$ and $n_\mathrm{cl}$, crossing the origin.
We also plot the regression function, Eq.~\eqref{main_eq_regression_function_001_001}, for DBSCAN and mean shift: $f_\mathrm{D} (x) = (0.0623392 \pm 0.0004168) x^{(0.999958 \pm 0.0007858)} + (0.0898713 \pm 0.07635)$ and $f_\mathrm{m} (x) = (1.49398 \pm 0.01079) x^{(0.528996 \pm 0.0007938)} + (- 4.63302 \pm 0.09461)$.
We applied the least mean square method to the mean values n the range of $n_\mathrm{ag} \in [10, 5000]$ to estimate the expected values and asymptotic standard error of parameters of this regression function.
Note that for small values of $n_\mathrm{ag}$, the estimate of $n_\mathrm{cl}$ may not be reliable because of the discreteness affecting the estimates.
In the case of DBSCAN, the estimated power, $c$ in Eq.~\eqref{main_eq_regression_function_001_001}, is almost unity; thus, the coefficient is a well-defined quantity in the thermodynamic limit.
Note that the $y$-intercept of the regression function, $b$ in Eq.~\eqref{main_eq_regression_function_001_001}, is almost zero.
On the other hand, the exponent is approximately $1/2$ and it implies that the coefficient is always meaninglessly trivial since $n_\mathrm{cl} / n_\mathrm{ag} \to 0$ in the thermodynamic limit.
In addition, the regression function, Eq.~\eqref{main_eq_regression_function_002_001}, for DBSCAN is $\tilde{f}_\mathrm{D} (x) = (0.0623314 \pm 1.287 \times 10^{-5}) x + (0.0718613 \pm 0.0372)$.
This estimated regression function supports that $\alpha = 0$ and $\beta > 0$, where $\alpha$ and $\beta$ are defined in Eq.~\eqref{main_eq_relation_n_cl_n_ag_001_001}.

\begin{figure}[t]
	\centering
	\includegraphics[scale=0.60]{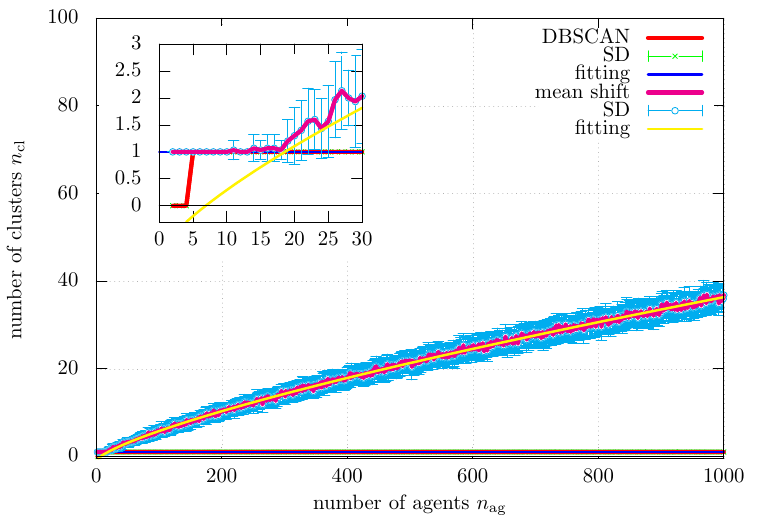}
	\caption{(Color online) Dependence of the expected value and standard deviation (SD) of $n_\mathrm{cl}$ on $n_\mathrm{ag}$. We set $r_\mathrm{V} = r_\mathrm{D} = 1.0$, $\eta = 4.0$, $\Delta t = 1.0$, $v_\mathrm{abs} = 0.030$, $n_\mathrm{min} = 5$, $t = 100$, and $n_\mathrm{rep} = 30$. We vary $L$ such that $n_\mathrm{ag} / L^2 = 4.0$. The regression functions, Eq.~\eqref{main_eq_regression_function_001_001}, for DBSCAN and mean shift are $f_\mathrm{D} (x) = (2.37045 \times 10^{-6} \pm 9.913 \times 10^{-6}) x^{(1.05079 \pm 0.5891)} + (0.999284 \pm 0.0007267)$ and $f_\mathrm{m} (x) = (0.215001 \pm 0.005851) x^{(0.746187 \pm 0.003709)} + (- 0.904683 \pm 0.08462)$, respectively. We used the data in the range of $n_\mathrm{ag} \in [10, 1000]$ to estimate the regression function.}
	\label{main_fig_fidelity_001_002_001_102}
\end{figure}

In Fig.~\ref{main_fig_fidelity_001_002_001_102}, we also show the dependence of the expected value and standard deviation (SD) of $n_\mathrm{cl}$ on $n_\mathrm{ag}$ in the case of $r_\mathrm{V} = r_\mathrm{D} = 1.0$.
We use the same values with Fig.~\ref{main_fig_fidelity_001_002_001_101} for the other parameters.
That is, we use parameters that belong to the white regime of Fig.~\ref{main_fig_fidelity_001_003_001_001} and Fig.~\ref{main_fig_fidelity_001_003_001_002}.
Figure~\ref{main_fig_fidelity_001_002_001_101} also clearly shows that $n_\mathrm{cl} = 1$ independently of $n_\mathrm{ag}$.
The regression functions, Eq.~\eqref{main_eq_regression_function_001_001}, for DBSCAN and mean shift are $f_\mathrm{D} (x) = (2.37045 \times 10^{-6} \pm 9.913 \times 10^{-6}) x^{(1.05079 \pm 0.5891)} + (0.999284 \pm 0.0007267)$ and $f_\mathrm{m} (x) = (0.215001 \pm 0.005851) x^{(0.746187 \pm 0.003709)} + (- 0.904683 \pm 0.08462)$, respectively.
We used the data in the range of $n_\mathrm{ag} \in [10, 1000]$ to estimate the regression function.
In addition, the regression function, Eq.~\eqref{main_eq_regression_function_002_001}, for DBSCAN is $\tilde{f}_\mathrm{D} (x) = (3.2392 \times 10^{-6} \pm 6.269 \times 10^{-7}) x + (0.999283 \pm 0.0003639)$.
This estimated regression function implies that $\alpha = 1$ and $\beta = 0$ for $\alpha$ and $\beta$ defined in Eq.~\eqref{main_eq_relation_n_cl_n_ag_001_001} since the smallness of the coefficient of the polynomial term, $a$ in Eq.~\eqref{main_eq_regression_function_001_001}, implies the redundancy of the functional form of Eq.~\eqref{main_eq_regression_function_001_001}.
Thus, we can see the discontinuous transition of $\alpha$ across the red and white regimes of Fig.~\ref{main_fig_fidelity_001_003_001_001} and Fig.~\ref{main_fig_fidelity_001_003_001_002}.
In Appendices~\ref{main_sec_additional_numerical_simulations_phase-diagram_001_001} and \ref{main_sec_additional_numerical_simulations_system-size-dependence_001_001}, we show additional numerical simulations on phase diagrams and system-size dependence, respectively.

\subsection{Discussions on phases A2 and A2'}

\begin{figure}[t]
	\centering
	\includegraphics[scale=0.70]{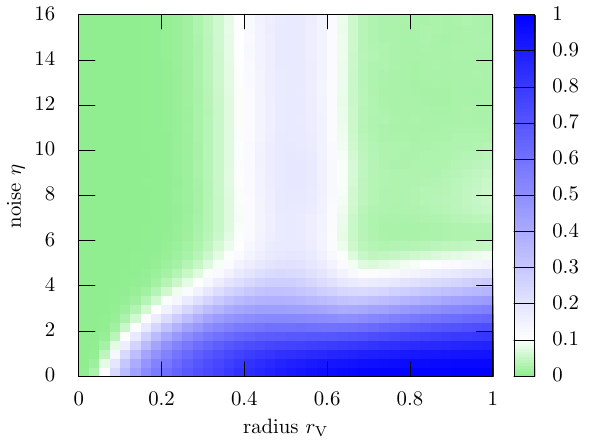}
	\caption{(Color online) Phase diagram of the intra-cluster order parameter, $v_\mathrm{icop} (t)$, in Eq.~\eqref{main_eq_intra-cluster_order_parameter_001_001}.}
	\label{main_fig_phase_diagram_order_parameter_001_001}
\end{figure}

We here discuss whether phases A2 and A2' are realized in the Vicsek model.
In Fig.~\ref{main_fig_phase_diagram_order_parameter_001_001}, we plot the intra-cluster order parameter, Eq.~\eqref{main_eq_intra-cluster_order_parameter_001_001}.
Figure~\ref{main_fig_phase_diagram_order_parameter_001_001} indicates that A2 is realized in the regime of $\eta \ge 3.0$ and $r_\mathrm{V} \in [0.3, 0.5]$ since the intra-cluster order parameter, Eq.~\eqref{main_eq_intra-cluster_order_parameter_001_001} has a nonzero value there.

\begin{figure}[t]
	\centering
	\includegraphics[scale=0.70]{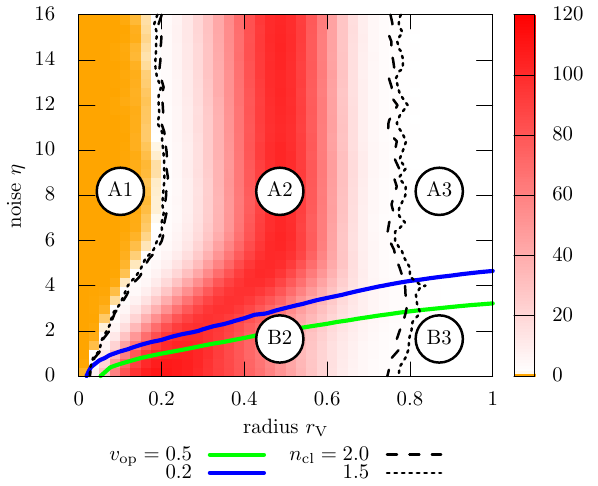}
	\caption{(Color online) Phase diagram of the Vicsek model with the labels of phases. The numerical setup is the same with Fig.~\ref{main_fig_fidelity_001_003_001_001} and the labels of the phases are defined in Fig.~\ref{main_schematic_DBSCAN_001_002}.}
	\label{main_fig_fidelity_101_003_001_001}
\end{figure}

In Fig.~\ref{main_fig_fidelity_101_003_001_001}, we added the labels defined in Fig.~\ref{main_schematic_DBSCAN_001_002} to Fig.~\ref{main_fig_fidelity_001_003_001_001}.
First, phase B1 does not exist in the phase diagram simply because $v_\mathrm{op} (t)$ in Eq.~\eqref{main_eq_order_parameter_Vicsek_001_001} cannot be nonzero without a global cluster or local clusters.
Second, Fig.~\ref{main_fig_fidelity_101_003_001_001} implies \textit{that phase A2' also does not exist in the phase diagram}.
Finally, the existence of phase A1 depends on $n_\mathrm{min}$ (the minimum number of nodes in any cluster) in DBSCAN as shown in Fig.~\ref{main_fig_fidelity_001_003_001_002}.

\section{Conclusions} \label{main_sec_conclusions_001_001}

In this paper, we investigate the cluster dynamics of the Vicsek model.
We first examine the relationship between the Vicsek model and DBSCAN, focusing on their cost functions.
We use DBSCAN, a widely used clustering algorithm, to estimate the cluster structure of the Vicsek model.
Our findings are important for two reasons.
Firstly, we identify new phases in the well-known Vicsek model, which has significant implications for natural science.
Mathematically, we have introduced two types of order parameters: the ratio of the number of clusters to that of agents and the mean velocity within clusters.
Secondly, we discover a mathematical relationship between the Vicsek model and DBSCAN.
The cost function of DBSCAN was not explicitly defined in the original paper, so this relationship was not previously understood.
Our work bridges two different fields and opens up new directions for research by combining them.

\begin{acknowledgments}
	We thank Kiyoshi Kanazawa, Hiroki Isobe, and Kyosuke Adachi for fruitful discussions.
	This work was supported by JSPS KAKENHI Grant Number JP25H01499.
\end{acknowledgments}

\appendix
\renewcommand{\thefigure}{\Alph{section}\arabic{figure}}

\setcounter{figure}{0}
\section{Table of variables} \label{main_sec_table_variables_001_001}

We have define a large number of variables to specify a system and algorithms; so we provide the table of variables in Table~\ref{main_table_fidelity_001_001}.
\begin{table*}[t]
	\caption{Definitions of variables.}
	\label{main_table_fidelity_001_001}
	\begin{ruledtabular}
		\begin{tabular}{cc}
			variable                        & description                                                             \\
			\hline
			$n_\mathrm{ag}$                 & number of agents                                                        \\
			$\bm{x}_i (t) \in \mathbb{R}^2$ & position of agent $i$ at time $t$                                       \\
			$\bm{v}_i (t) \in \mathbb{R}^2$ & velocity of agent $i$ at time $t$                                       \\
			$\theta_i (t) \in \mathbb{R}$   & direction of agent $i$ at time $t$                                      \\
			$v_\mathrm{abs}$                & absolute value of the velocity of all the agents                        \\
			$r_\mathrm{V}$                  & range of the interaction of the Vicsek model                            \\
			$r_\mathrm{D}$                  & range of neighbors of DBSCAN                                            \\
			$N_{i, r_\mathrm{V}} (t)$       & \# of neighbors of agent $i$ within distance $r_\mathrm{V}$ at time $t$ \\
			$v_\mathrm{op} (t)$             & order parameter originally proposed by Vicsek \textit{et al}.           \\
			$v_\mathrm{icop} (t)$           & intra-cluster order parameter                                           \\
			$\eta$                          & noise magnitude ($\Xi_i (t) \sim [- \eta / 2, \eta / 2]$)               \\
			$n_\mathrm{min}$                & minimum number of agents that each cluster must have in it for DBSCAN
		\end{tabular}
	\end{ruledtabular}
\end{table*}

\setcounter{figure}{0}
\section{Additional numerical simulations (phase diagrams)} \label{main_sec_additional_numerical_simulations_phase-diagram_001_001}

\begin{figure}[t]
	\centering
	\includegraphics[scale=0.70]{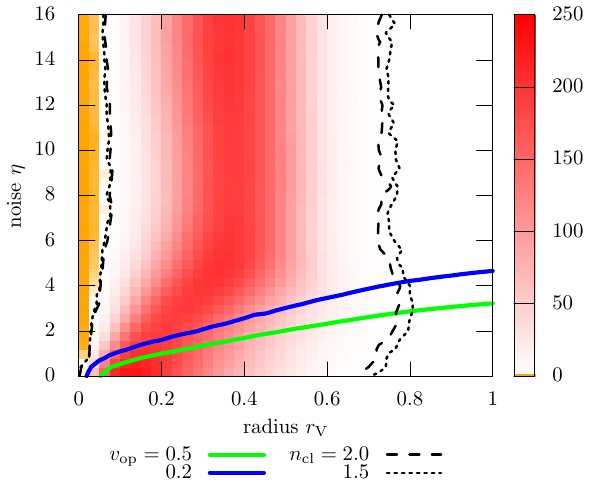}
	\caption{(Color online) Dependence of $n_\mathrm{cl}$ (heat map) and $v_\mathrm{op} (t)$ (solid lines) on $r_\mathrm{V}$ (horizontal) and $\eta$ (vertical). We set $r_\mathrm{V} = r_\mathrm{D}$, $n_\mathrm{ag} = 1600$, $L = 20.0$, $\Delta t = 1.0$, $v_\mathrm{abs} = 0.030$, $n_\mathrm{min} = 3$, $t = 100$, and $n_\mathrm{rep} = 30$. The black dotted and dashed lines are contour lines for the number of estimated number of clusters.}
	\label{main_fig_fidelity_001_003_002_001}
\end{figure}

\begin{figure}[t]
	\centering
	\includegraphics[scale=0.70]{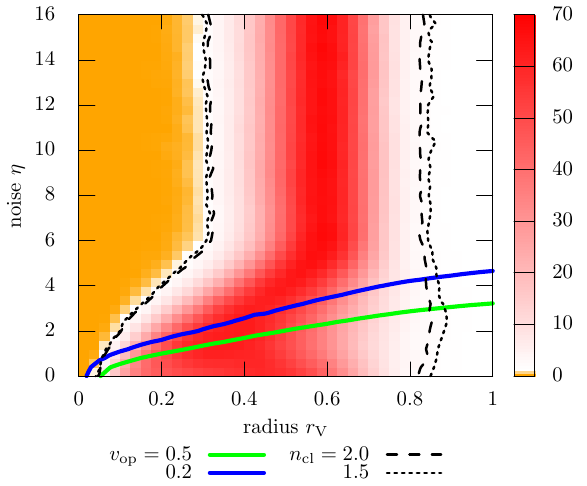}
	\caption{(Color online) Dependence of $n_\mathrm{cl}$ (heat map) and $v_\mathrm{op} (t)$ (solid lines) on $r_\mathrm{V}$ (horizontal) and $\eta$ (vertical). We set $r_\mathrm{V} = r_\mathrm{D}$, $n_\mathrm{ag} = 1600$, $L = 20.0$, $\Delta t = 1.0$, $v_\mathrm{abs} = 0.030$, $n_\mathrm{min} = 7$, $t = 100$, and $n_\mathrm{rep} = 30$. The black dotted and dashed lines are contour lines for the number of estimated number of clusters.}
	\label{main_fig_fidelity_001_003_002_002}
\end{figure}

\begin{figure}[t]
	\centering
	\includegraphics[scale=0.70]{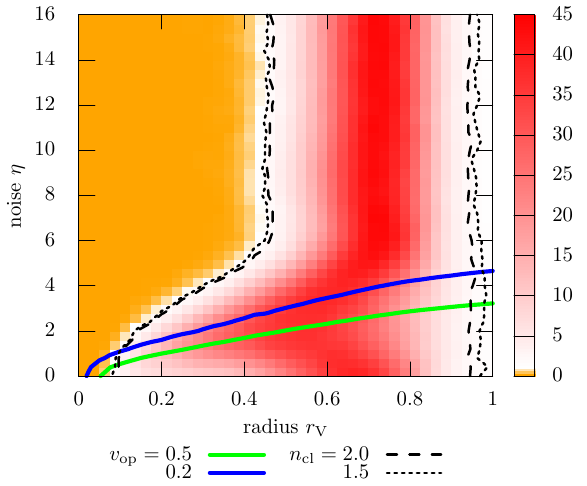}
	\caption{(Color online) Dependence of $n_\mathrm{cl}$ (heat map) and $v_\mathrm{op} (t)$ (solid lines) on $r_\mathrm{V}$ (horizontal) and $\eta$ (vertical). We set $r_\mathrm{V} = r_\mathrm{D}$, $n_\mathrm{ag} = 1600$, $L = 20.0$, $\Delta t = 1.0$, $v_\mathrm{abs} = 0.030$, $n_\mathrm{min} = 10$, $t = 100$, and $n_\mathrm{rep} = 30$. The black dotted and dashed lines are contour lines for the number of estimated number of clusters.}
	\label{main_fig_fidelity_001_003_002_003}
\end{figure}

In Fig.~\ref{main_fig_fidelity_001_003_002_001}, we plot the dependence of $n_\mathrm{cl}$ and $v_\mathrm{op} (t)$ on $r_\mathrm{V}$ and $\eta$.
We set $r_\mathrm{V} = r_\mathrm{D}$, $n_\mathrm{ag} = 1600$, $L = 20.0$, $\Delta t = 1.0$, $v_\mathrm{abs} = 0.030$, $n_\mathrm{min} = 3$, and $t = 100$.
We repeated the same calculations 30 times ($n_\mathrm{rep} = 30$) to compute the statistical means.
In Fig.~\ref{main_fig_fidelity_001_003_002_002}, we plot the dependence of $n_\mathrm{cl}$ and $v_\mathrm{op} (t)$ on $r_\mathrm{V}$ and $\eta$ for the Vicsek model in the case of $n_\mathrm{min} = 7$.
In Fig.~\ref{main_fig_fidelity_001_003_002_003}, we also depict the dependence of $n_\mathrm{cl}$ and $v_\mathrm{op} (t)$ on $r_\mathrm{V}$ and $\eta$ for the Vicsek model in the case of $n_\mathrm{min} = 10$.

\setcounter{figure}{0}
\section{Additional numerical simulations (system-size dependence)} \label{main_sec_additional_numerical_simulations_system-size-dependence_001_001}

We show additional numerical simulations on the dependence of the expected value and standard deviation (SD) of $n_\mathrm{cl}$ on $n_\mathrm{ag}$ for the Vicsek model near the boundary between $(\alpha = 0, \beta > 0)$ and $(\alpha = 1.0, \beta = 0)$.

\begin{figure}[t]
	\centering
	\includegraphics[scale=0.60]{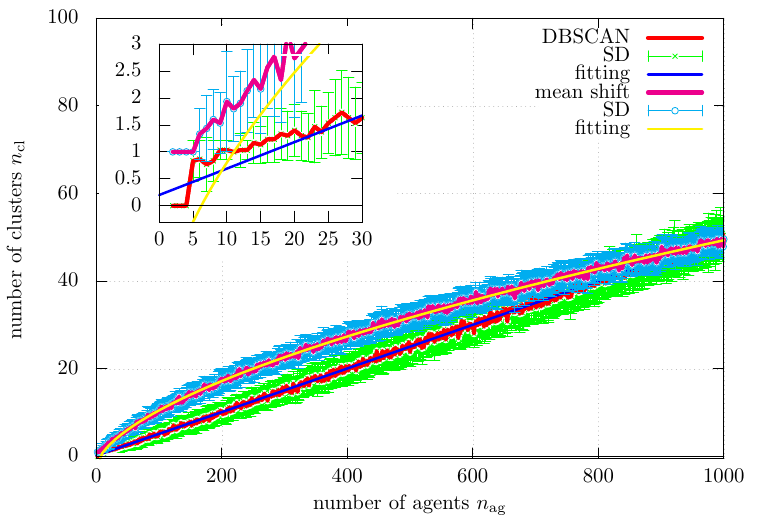}
	\caption{(Color online) Dependence of the expected value and standard deviation (SD) of $n_\mathrm{cl}$ on $n_\mathrm{ag}$. We set $r_\mathrm{V} = r_\mathrm{D} = 0.5$, $\eta = 4.0$, $\Delta t = 1.0$, $v_\mathrm{abs} = 0.030$, $n_\mathrm{rep} = 30$, $n_\mathrm{min} = 5$, and $t = 100$. We vary $L$ such that $n_\mathrm{ag} / L^2 = 4.0$. The regression functions, Eq.~\eqref{main_eq_regression_function_001_001}, for DBSCAN and mean shift are $f_\mathrm{D} (x) = (0.0484874 \pm 0.001447) x^{(1.00388 \pm 0.004189)} + (0.20035 \pm 0.08191)$ and $f_\mathrm{m} (x) = (0.803264 \pm 0.01997) x^{(0.602986 \pm 0.003285)} + (- 2.41772 \pm 0.1426)$, respectively. We used the data in regime of $n_\mathrm{ag} \in [10, 1000]$ to estimate the linear regression function.}
	\label{main_fig_fidelity_001_002_002_103}
\end{figure}

In Fig.~\ref{main_fig_fidelity_001_002_002_103}, we plot the dependence of the expected value and standard deviation (SD) of $n_\mathrm{cl}$ on $n_\mathrm{ag}$.
We set $r_\mathrm{V} = r_\mathrm{D} = 0.5$, $\eta = 4.0$, $\Delta t = 1.0$, $v_\mathrm{abs} = 0.030$, $n_\mathrm{rep} = 30$, $n_\mathrm{min} = 5$, and $t = 100$.
We vary $L$ such that $n_\mathrm{ag} / L^2 = 4.0$.
The regression functions, Eq.~\eqref{main_eq_regression_function_001_001}, for DBSCAN and mean shift are $f_\mathrm{D} (x) = (0.0484874 \pm 0.001447) x^{(1.00388 \pm 0.004189)} + (0.20035 \pm 0.08191)$ and $f_\mathrm{m} (x) = (0.803264 \pm 0.01997) x^{(0.602986 \pm 0.003285)} + (- 2.41772 \pm 0.1426)$, respectively.
In addition, the regression function, Eq.~\eqref{main_eq_regression_function_002_001}, for DBSCAN is $\tilde{f}_\mathrm{D} (x) = (0.0498399 \pm 6.723 \times 10^{-5}) x + (0.135442 \pm 0.03902)$.

\begin{figure}[t]
	\centering
	\includegraphics[scale=0.60]{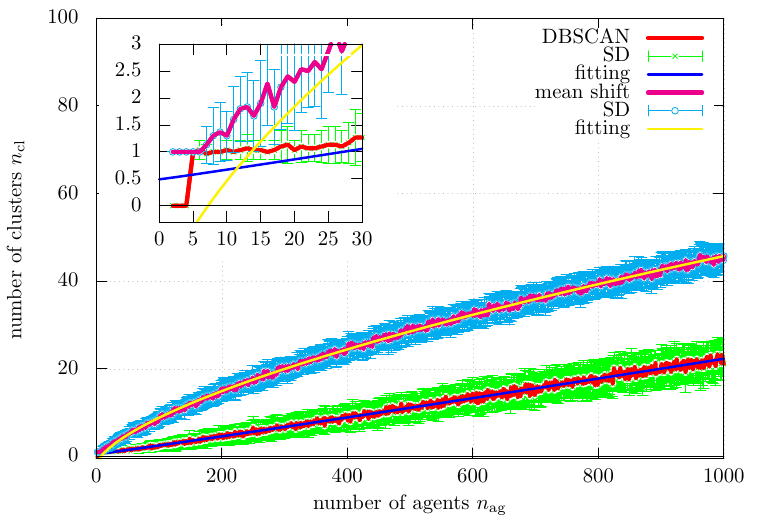}
	\caption{(Color online) Dependence of the expected value and standard deviation (SD) of $n_\mathrm{cl}$ on $n_\mathrm{ag}$. We set $r_\mathrm{V} = r_\mathrm{D} = 0.6$, $\eta = 4.0$, $\Delta t = 1.0$, $v_\mathrm{abs} = 0.030$, $n_\mathrm{rep} = 30$, $n_\mathrm{min} = 5$, and $t = 100$. We vary $L$ such that $n_\mathrm{ag} / L^2 = 4.0$. The regression functions, Eq.~\eqref{main_eq_regression_function_001_001}, for DBSCAN and mean shift are $f_\mathrm{D} (x) = (0.0163892 \pm 0.0009795) x^{(1.04122 \pm 0.008413)} + (0.490484 \pm 0.06811)$ and $f_\mathrm{m} (x) = (0.01395 \pm 0.06811) x^{(0.645449 \pm 0.003376)} + (- 1.97723 \pm 0.1222)$, respectively. We used the data in regime of $n_\mathrm{ag} \in [10, 1000]$ to estimate the linear regression function.}
	\label{main_fig_fidelity_001_002_002_101}
\end{figure}

In Fig.~\ref{main_fig_fidelity_001_002_002_101}, we plot the dependence of the expected value and standard deviation (SD) of $n_\mathrm{cl}$ on $n_\mathrm{ag}$ for the Vicsek model in the case of $r_\mathrm{V} = r_\mathrm{D} = 0.6$.
The regression functions, Eq.~\eqref{main_eq_regression_function_001_001}, for DBSCAN and mean shift are $f_\mathrm{D} (x) = (0.0163892 \pm 0.0009795) x^{(1.04122 \pm 0.008413)} + (0.490484 \pm 0.06811)$ and $f_\mathrm{m} (x) = (0.01395 \pm 0.06811) x^{(0.645449 \pm 0.003376)} + (- 1.97723 \pm 0.1222)$, respectively.
In addition, the regression function, Eq.~\eqref{main_eq_regression_function_002_001}, for DBSCAN is $\tilde{f}_\mathrm{D} (x) = (0.0219426 \pm 5.888 \times 10^{-5}) x + (0.190476 \pm 0.03418)$.
In this setup, $n_\mathrm{cl}$ increases linearly with $n_\mathrm{ag}$ though $n_\mathrm{cl}$ shows a step function-like behavior on $n_\mathrm{ag}$ for small $n_\mathrm{ag}$.

\begin{figure}[t]
	\centering
	\includegraphics[scale=0.60]{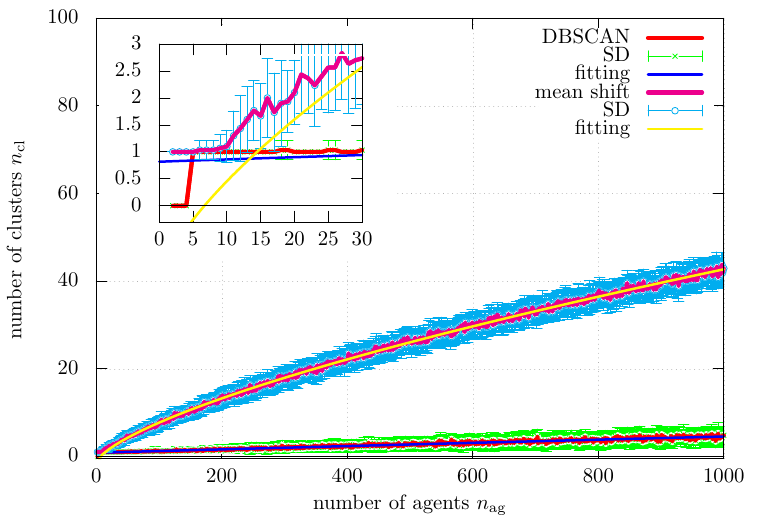}
	\caption{(Color online) Dependence of the expected value and standard deviation (SD) of $n_\mathrm{cl}$ on $n_\mathrm{ag}$. We set $r_\mathrm{V} = r_\mathrm{D} = 0.7$, $\eta = 4.0$, $\Delta t = 1.0$, $v_\mathrm{abs} = 0.030$, $n_\mathrm{rep} = 30$, $n_\mathrm{min} = 5$, and $t = 100$. We vary $L$ such that $n_\mathrm{ag} / L^2 = 4.0$. The regression functions, Eq.~\eqref{main_eq_regression_function_001_001}, for DBSCAN and mean shift are $f_\mathrm{D} (x) = (0.00458696 \pm 0.0007248) x^{(0.969634 \pm 0.02212)} + (0.817777 \pm 0.03404)$ and $f_\mathrm{m} (x) = (0.389796 \pm 0.01045) x^{(0.684613 \pm 0.003613)} + (- 1.42979 \pm 0.111)$, respectively. We used the data in the regime of $n_\mathrm{ag} \in [10, 1000]$ to estimate the linear regression function.}
	\label{main_fig_fidelity_001_002_002_102}
\end{figure}

In Fig.~\ref{main_fig_fidelity_001_002_002_102}, we plot the dependence of the expected value and standard deviation (SD) of $n_\mathrm{cl}$ on $n_\mathrm{ag}$ for the Vicsek model in the case of $r_\mathrm{V} = r_\mathrm{D} = 0.7$.
The regression functions, Eq.~\eqref{main_eq_regression_function_001_001}, for DBSCAN and mean shift are $f_\mathrm{D} (x) = (0.00458696 \pm 0.0007248) x^{(0.969634 \pm 0.02212)} + (0.817777 \pm 0.03404)$ and $f_\mathrm{m} (x) = (0.389796 \pm 0.01045) x^{(0.684613 \pm 0.003613)} + (- 1.42979 \pm 0.111)$, respectively.
In addition, the regression function, Eq.~\eqref{main_eq_regression_function_002_001}, for DBSCAN is $\tilde{f}_\mathrm{D} (x) = (0.00370084 \pm 2.688 \times 10^{-5}) x + (0.857951 \pm 0.0156)$.
In this setup, $(\alpha \approx 1.0, \beta > 0)$ is realized.
However, this parameter regime is quite small.
Further investigation is necessary to verify the robustness of this phase in the thermodynamic limit.

\begin{figure}[t]
	\centering
	\includegraphics[scale=0.60]{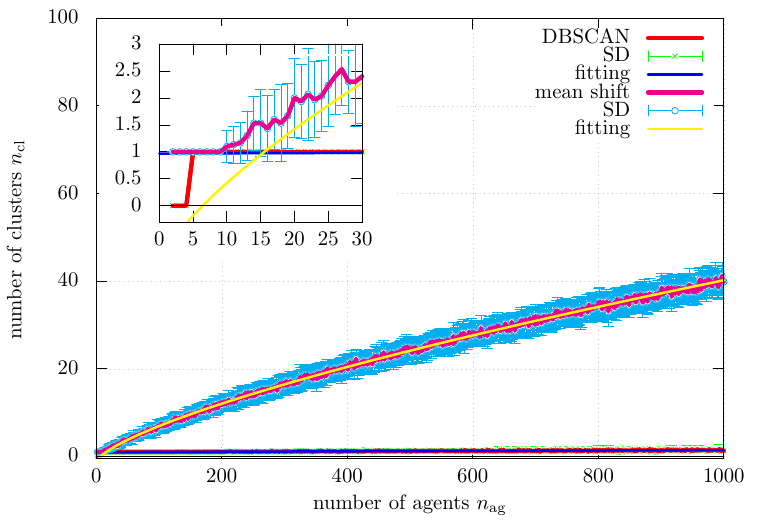}
	\caption{(Color online) Dependence of the expected value and standard deviation (SD) of $n_\mathrm{cl}$ on $n_\mathrm{ag}$. We set $r_\mathrm{V} = r_\mathrm{D} = 0.8$, $\eta = 4.0$, $\Delta t = 1.0$, $v_\mathrm{abs} = 0.030$, $n_\mathrm{rep} = 30$, $n_\mathrm{min} = 5$, and $t = 100$. We vary $L$ such that $n_\mathrm{ag} / L^2 = 4.0$. The regression functions, Eq.~\eqref{main_eq_regression_function_001_001}, for DBSCAN and mean shift are $f_\mathrm{D} (x) = (0.000520394 \pm 0.0002336) x^{(0.980815 \pm 0.0629)} + (0.968394 \pm 0.01166)$ and $f_\mathrm{m} (x) = (0.305275 \pm 0.008053) x^{(0.710531 \pm 0.003573)} + (- 1.14446 \pm 0.09729)$, respectively. We used the data in regime of $n_\mathrm{ag} \in [10, 1000]$ to estimate the linear regression function.}
	\label{main_fig_fidelity_001_002_002_104}
\end{figure}

In Fig.~\ref{main_fig_fidelity_001_002_002_104}, we plot the dependence of the expected value and standard deviation (SD) of $n_\mathrm{cl}$ on $n_\mathrm{ag}$ for the Vicsek model in the case of $r_\mathrm{V} = r_\mathrm{D} = 0.8$.
The regression functions, Eq.~\eqref{main_eq_regression_function_001_001}, for DBSCAN and mean shift are $f_\mathrm{D} (x) = (0.000520394 \pm 0.0002336) x^{(0.980815 \pm 0.0629)} + (0.968394 \pm 0.01166)$ and $f_\mathrm{m} (x) = (0.305275 \pm 0.008053) x^{(0.710531 \pm 0.003573)} + (- 1.14446 \pm 0.09729)$, respectively.
In addition, the regression function, Eq.~\eqref{main_eq_regression_function_002_001}, for DBSCAN is $\tilde{f}_\mathrm{D} (x) = (0.000454206 \pm 9.318 \times 10^{-6}) x + (0.971467 \pm 0.005408)$.

\begin{figure}[t]
	\centering
	\includegraphics[scale=0.60]{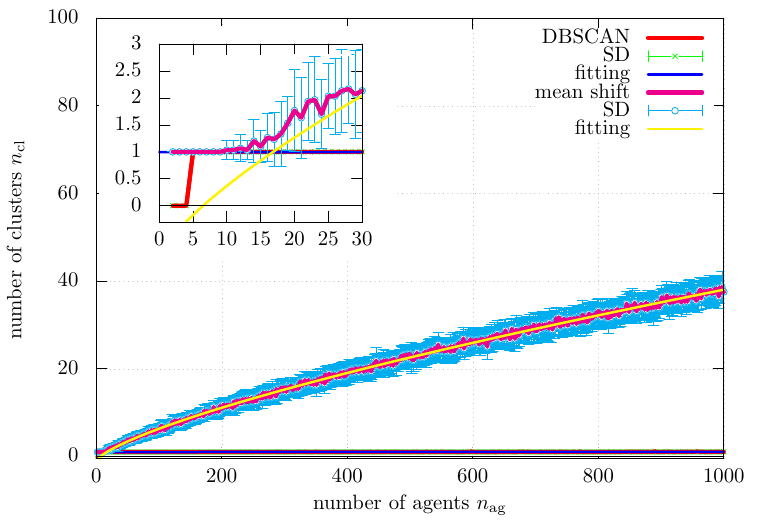}
	\caption{(Color online) Dependence of the expected value and standard deviation (SD) of $n_\mathrm{cl}$ on $n_\mathrm{ag}$. We set $r_\mathrm{V} = r_\mathrm{D} = 0.9$, $\eta = 4.0$, $\Delta t = 1.0$, $v_\mathrm{abs} = 0.030$, $n_\mathrm{rep} = 30$, $n_\mathrm{min} = 5$, and $t = 100$. We vary $L$ such that $n_\mathrm{ag} / L^2 = 4.0$. The regression functions, Eq.~\eqref{main_eq_regression_function_001_001}, for DBSCAN and mean shift are $f_\mathrm{D} (x) = (2.48221 \times 10^{-5} \pm 3.606 \time 10^{-5}) x^{(1.0907 \pm 0.2052)} + (0.996329 \pm 0.0033)$ and $f_\mathrm{m} (x) = (0.255057 \pm 0.006794) x^{(0.728259 \pm 0.003619)} + (- 0.994408 \pm 0.08973)$, respectively. We used the data in regime of $n_\mathrm{ag} \in [10, 1000]$ to estimate the linear regression function.}
	\label{main_fig_fidelity_001_002_002_105}
\end{figure}

In Fig.~\ref{main_fig_fidelity_001_002_002_105}, we plot the dependence of the expected value and standard deviation (SD) of $n_\mathrm{cl}$ on $n_\mathrm{ag}$ for the Vicsek model in the case of $r_\mathrm{V} = r_\mathrm{D} = 1.0$.
The regression functions, Eq.~\eqref{main_eq_regression_function_001_001}, for DBSCAN and mean shift are $f_\mathrm{D} (x) = (2.48221 \times 10^{-5} \pm 3.606 \time 10^{-5}) x^{(1.0907 \pm 0.2052)} + (0.996329 \pm 0.0033)$ and $f_\mathrm{m} (x) = (0.255057 \pm 0.006794) x^{(0.728259 \pm 0.003619)} + (- 0.994408 \pm 0.08973)$, respectively.
In addition, the regression function, Eq.~\eqref{main_eq_regression_function_002_001}, for DBSCAN is $\tilde{f}_\mathrm{D} (x) = (4.70149 \times 10^{-5} \pm 2.965 \times 10^{-6}) x + (0.995026 \pm 0.001721)$.

\bibliography{paper_Vicsek_DBSCAN_000_001_main}

\end{document}